\documentclass[%
 aip,
 amsmath,amssymb,
 reprint,%
]{revtex4-1}

\usepackage{graphicx}
\usepackage{dcolumn}
\usepackage{bm}
\usepackage[mathlines]{lineno}

\usepackage{booktabs}

\usepackage[utf8]{inputenc}
\usepackage[T1]{fontenc}
\usepackage{mathptmx}
\usepackage{etoolbox}
\usepackage{color}
\usepackage{soul}
\soulregister\cite7
\soulregister\ref7
\makeatletter
\def\@email#1#2{%
 \endgroup
 \patchcmd{\titleblock@produce}
  {\frontmatter@RRAPformat}
  {\frontmatter@RRAPformat{\produce@RRAP{*#1\href{mailto:#2}{#2}}}\frontmatter@RRAPformat}
  {}{}
}%
\makeatother

\begin{document}

\preprint{AIP/123-QED}

\title{Realizing Laser-driven Deuteron Acceleration with Low Energy Spread via \textit{In-situ} D$_2$O-deposited Target}

\author{Tianyun Wei}
\affiliation{%
Institute of Laser Engineering, Osaka University, Suita 567-0871, Japan
}%

\author{Yasunobu Arikawa}
\affiliation{%
Institute of Laser Engineering, Osaka University, Suita 567-0871, Japan
}%

\author{Seyed Reza Mirfayzi}
\affiliation{%
Tokamak Energy ltd, 173 Brook Dr, Milton, Abingdon OX14 4SD, United Kingdom
}%

\author{Yanjun Gu}
\affiliation{%
SANKEN (Institute of Scientific and Industrial Research), Osaka University,  Ibaraki 567-0047, Japan
}%

\author{Takehito Hayakawa}
\affiliation{%
Kansai Institute for Photon Science, National Institutes for Quantum Science and Technology (QST), 8-1-7 Umemidai, Kizugawa, Kyoto 619-0215, Japan
}%
\affiliation{%
Institute of Laser Engineering, Osaka University, Suita 567-0871, Japan
}%

\author{Alessio Morace}
\affiliation{%
Institute of Laser Engineering, Osaka University, Suita 567-0871, Japan
}%

\author{Kunioki Mima}
\affiliation{%
Institute of Laser Engineering, Osaka University, Suita 567-0871, Japan
}%

\author{Zechen Lan}
\affiliation{%
Institute of Laser Engineering, Osaka University, Suita 567-0871, Japan
}%

\author{Ryuya Yamada}
\affiliation{%
Institute of Laser Engineering, Osaka University, Suita 567-0871, Japan
}%

\author{Kohei Yamanoi}
\affiliation{%
Institute of Laser Engineering, Osaka University, Suita 567-0871, Japan
}%

\author{Koichi Honda}
\affiliation{%
Institute of Laser Engineering, Osaka University, Suita 567-0871, Japan
}%

\author{ Sergei V. Bulanov}
\affiliation{%
Extreme Light InfrastructureI ERIC, ELI–Beamlines Facility, Za Radnici 835, Dolni Brezany 25241, Czech Republic
}%
\affiliation{%
Kansai Institute for Photon Science, National Institutes for Quantum Science and Technology (QST), 8-1-7 Umemidai, Kizugawa, Kyoto 619-0215, Japan
}%

\author{Akifumi Yogo}%
\email{yogo-a@ile.osaka-u.ac.jp}
\affiliation{%
Institute of Laser Engineering, Osaka University, Suita 567-0871, Japan
}%

\date{\today}

\begin{abstract}

Generation of quasi-monoenergetic ion pulse by laser-driven acceleration is one of the hot topics in laser plasma physics.
In this study, we present a new method for the \textit{In-situ} deposition of an ultra-thin D$_2$O layer on the surface of an aluminum foil target utilizing a spherical D$_2$O capsule. 
Employing a 10$^{19}$~W/cm$^2$ laser, we achieve the acceleration of 10.8~MeV deuterons with an energy spread of $\Delta$E/E = 4.6\% in the most favorable shot. 
The energy spread depends on  the exposure time of the D$_2$O capsule in the vacuum chamber.
This method has the potential to extend its applicability to other ion species.

\end{abstract}

\maketitle

\section{Introduction}

The progress in ultra-intense laser systems\cite{danson2019petawatt} has opened the way for numerous emerging fields, notably laser-driven ion acceleration \cite{macchi2013ion} capable of reaching energies in the tens of MeV range \cite{higginson2018near, yogo2017boosting}. 
These advancements have enabled applications across scientific research, industry, and health sectors including compact neutron sources\cite{yogo2023laser}, proton radiography \cite{romagnani2008proton}, cancer therapy \cite{bulanov2002feasibility,ledingham2014towards,bulanov2014laser}, and fast ignition initiatives\cite{roth2001fast,fernandez2009progress}. 
The current challenges include maximizing energy enhancement \cite{higginson2018near,alejo2017high} and reducing energy bandwidth\cite{esirkepov2002proposed,maksimchuk2004high,schwoerer2006laser,hegelich2006laser,ter2006quasimonoenergetic, henig2009radiation, ramakrishna2010laser, jung2011monoenergetic,kar2012ion, palaniyappan2015efficient,pak2018collisionless,scott2018dual,hilz2018isolated,li2019laser, bagchi2021quasi,ahmed2021high}, 
which are crucial considerations for applications such as laser-based ion cancer therapy.

Target normal sheath acceleration (TNSA)\cite{wilks2001energetic,passoni2010target} is one of the most well-established mechanisms due to its low requirements both in target and laser conditions, making it applicable for almost all high power laser facilities with intensities above 10$^{18}$~W/cm$^2$. 
However, the TNSA accelerated ions typically feature a continuous energy distribution because the ions are accelerated by the sheath field created by hot electrons. 
On the other hand, radiation pressure acceleration (RPA)\cite{esirkepov2004highly} and collisionless shock acceleration (CSA)\cite{silva2004proton} are anticipated to achieve mono-energetic ion acceleration although they require high intensity of 10$^{20}$~W/cm$^2$.
The current experimental conditions impose limitations due to the ultra-high laser intensity and low-density target requirements and have proved to be challenging.
Thus, quasi-mono energetic ion acceleration via the TNSA mechanism emerges as a preferable approach feasible for most laser facilities.

In the TNSA mechanism, ions such as protons are generally accelerated from a contamination layer on the surface of a target. Therefore, the control of the contamination layer\cite{macchi2013ion} through structural modifications \cite{schwoerer2006laser} or thickness adjustments \cite{hegelich2006laser} is crucial for achieving quasi-monoenergetic ion acceleration via TNSA mechanism.
In addition, various methods for thin layer targets for high-frequency laser have been developed.\cite{prencipe2017targets}
B. M. Hegelich et al. \cite{hegelich2006laser} reported generation of quasi-monoenergetic carbon ions with an energy spread of $\Delta $E/E $\simeq$ 17\% using the TNSA mechanism. 
This was achieved by heating the metal target to eliminate hydrogen contamination to leave only an ultra-thin layer of carbon.
It is reported that a new layer of D$_2$O  on a target could be generated by injecting heavy
water into a chamber\cite{hou2011laser}, so that deuteron acceleration was enhanced  although its energy spectrum was broad.

In this study, we refine this approach by placing a heavy water capsule adjacent to an aluminum target within the laser chamber under vacuum condition at room temperature. 
The D$_2$O molecules are released from the spherical capsule under the vacuum environment and subsequently accumulate on the surface of the aluminum target, resulting in the formation of an ultra-thin layer of D$_2$O.
We conduct ion acceleration experiments with the D$_2$O deposited Al targets using a high-power laser and provide the energy spectra of the accelerated ions measured with a Thomson parabola ion spectrometer (TPIS)\cite{alejo2016high,tosaki2017evaluation}. 
Quasi-monoenergetic deuterons with energies close to 11~MeV with an energy spread ($\Delta $E/E) of approximately 4.6\% are measured in the most favorable shot.

 \section{Experiment}

\begin{figure*}

\includegraphics[width=15cm]{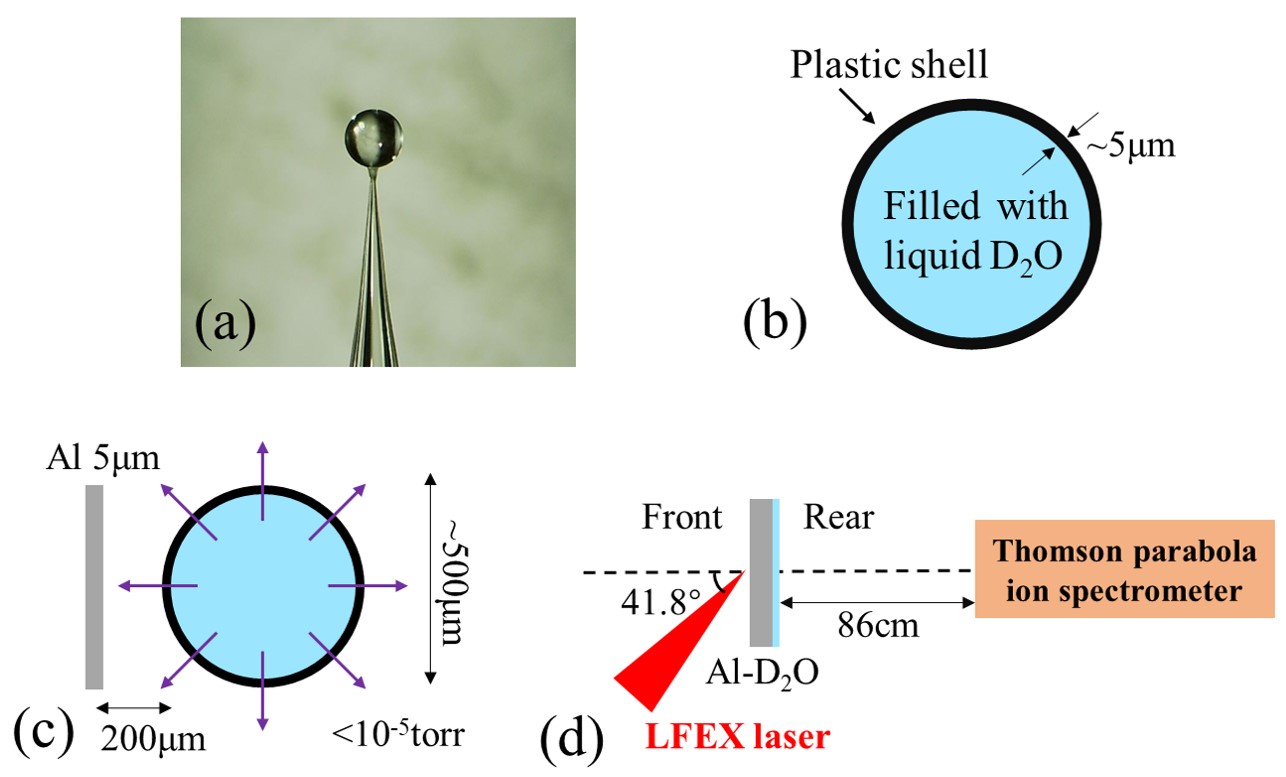}
\caption{ (a) Photograph of the D$_2$O capsule. (b) Structure of the D$_2$O capsule. (c) Schematic view of accumulation of  D$_2$O molecules leaked from the capsule on the surface of Al target in the vacuum chamber. (d) Ion acceleration experiment setup with the Al target and the ultra-thin D$_2$O layer using the LFEX laser.}
\label{fEXPSET}

\end{figure*}

In the experiments, a heavy water capsule is used to deposit an ultra-thin D$_2$O layer on an aluminum target. 
The photograph and the structure of the D$_2$O capsule are shown in Figs.~\ref{fEXPSET}(a) and (b). 
The capsule with a plastic shell containing liquid D$_2$O is fabricated with water-in-oil-in-water (W/O/W) double emulsions methods \cite{takagi1991development,iwasa2018controlled}.
It has a diameter of approximately 500~$\mu$m with a shell thickness of $\sim$5~$\mu$m. Consequently, the capsule typically holds approximately 0.06~$\mu$l of liquid D$_2$O.
To deposit a D$_2$O layer onto an aluminum target, the D$_2$O capsule is placed 200-$\mu$m away from the target in the laser irradiation chamber and is exposed in a low pressure of $< 10^{-5}$~ torr as shown in Fig.\ref{fEXPSET}(c).
It is important to note that under such low pressure, the liquid D$_2$O leaks out from the capsule in form of D$_2$O molecules due to the nanometer-scale porous structure\cite{davis2002ordered} of the plastic shell. 
The aluminum target can then capture some of the D$_2$O molecules, resulting in the formation of an ultra-thin layer of D$_2$O on the surface of the target.
Furthermore, the thickness of the D$_2$O layer is expected to increase proportionally with the duration time that the capsule is placed near the target.

The ion acceleration experiments are conducted utilizing the petawatt laser system LFEX\cite{kawanaka20083} at the Institute of Laser Engineering in Osaka University. 
The typical parameters of the laser used in the experiments are an energy of approximately 600~J on the target, a pulse duration of approximately 1.5~ps, a spot size of $\sim$50~$\mu$m, and an intensity of approximately $1\times10^{19}$W/cm$^2$.
The experiment setup with the Al-D$_2$O target is shown in Fig.~\ref{fEXPSET}(d).  
The LFEX laser irradiates on the front side of the target with 41.8$^\circ$ away from the normal direction of the target.
The Thomson parabola ion spectrometer (TPIS) is set at the target normal direction in the rear side to measure the energy spectra of the accelerated ions. 
The heavy ions including carbon, oxygen, and aluminum are stopped by an Al filter with a thickness of 100~${\mu}$m located on the front of an imaging plate (IP) of the TPIS\cite{golovin2021calibration}, and thus only light ions of protons and deuterons are measured.
Because the Al filter is directly attached to the IP, the ion energies are not affected significantly by the filter. Although the energy resolution is slightly expanded, we evaluated the energy spectra with a resolution of 0.2~MeV which is higher than this effect.

\begin{figure*}

\includegraphics[width=15cm]{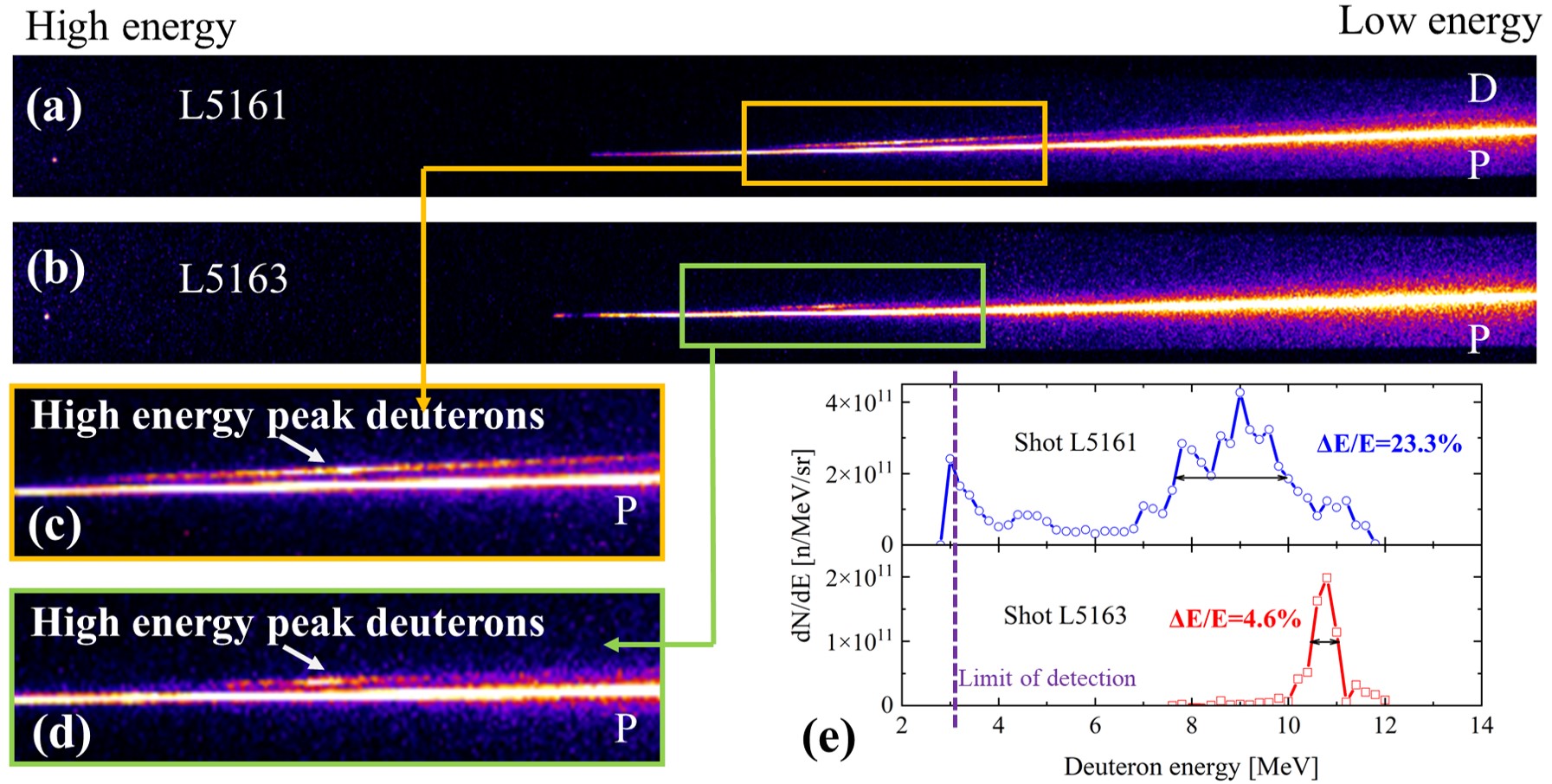}
\caption{(a) Image measured with the TPIS for shot L5161. D$_2$O depositing time for accumulation of a  D$_2$O layer on the target is $\sim$100~min. (b) Image for shot L5163. D$_2$O depositing time is $\sim$40~min. (c) Expanded image of (a). (d) Expanded image of (b). High-energy deuteron peaks are observed in the both shots. (e) Experiment deuteron energy spectra for the both shots. Quasi-mono energetic deuterons are measured, and the energy spread $\Delta $E/E becomes smaller for shorter D$_2$O depositing time}
\label{fEXPRESULT}

\end{figure*}

Figs.~\ref{fEXPRESULT}(a) and (b) show the images measured by the TPIS for two laser shots of L5161 and L5163.
In the two shots, we change the exposure time that the D$_2$O capsule is located near the target
to change the thickness of the D$_2$O accumulation layer.
The exposure time are 100~min for L5161 and  40~min for L5163. Figs.~\ref{fEXPRESULT}(a) and (b) show the strong proton signals and relatively weak deuteron signals.
In the high energy area, the deuteron signals are clearly observed and separated from the proton signals [Figs.~\ref{fEXPRESULT}(c) and (d)].
The deuteron energy spectra analyzed from the TPIS data is shown in Fig.~\ref{fEXPRESULT}(e), 
where the quasi-monoenergetic deuteron peaks appear near 10~MeV in the both shots.
It is considered that the thickness of the D$_2$O layer decreases with decreasing the exposure time of the D$_2$O capsule in the vacuum chamber.
Thus, the D$_2$O layer thickness for L5163 is expected to be shorter than that for L5161.
The measured energy spectra show that as the D$_2$O layer becomes thinner
the energy spread $\Delta$E/E decreases from 23.3\% to 4.6\%
and the peak energy slightly increases from 9.0~MeV to 10.8~MeV.
The deuteron flux at the peak energy, however, decreases from 4.2$\times 10^{11}$~n/MeV/sr to 2.0$\times 10^{11}$~n/MeV/sr when the thickness of the D$_2$O layer decreases. 
To estimate the evaporation rate of D$_2$O molecules from the capsule and the thickness of the accumulated D$_2$O layer we measure the D$_2$O gas partial pressure in another vacuum chamber using a quadrupole mass spectrometer, but
the difference between the D$_2$O gas partial pressures with/without the D$_2$O capsule is lower than the detection limit.
It was reported\cite{alfonso2006using} that the half-life of the leakage rate of 5.4~atm deuterium gas from an inertial confinement fusion target under vacuum is 15-40 days. We estimate the thickness of the D$_2$O layer accumulated under our experimental condition using the half-life of 15 day. The estimated thicknesses are 82~nm for L5161 and 33~nm for L5163.
As presented later, 
a particle-in-cell (PIC) simulation results simulation result suggests that the thickness of the D$_2$O layer is in order of several tens of nm.

Figure 3 shows the deuteron energy spectra for the L4694 and L4701 shots. The laser irradiation condition is almost identical with those for L5161 and L5163. In the L4694 and L4701 shots, the D$_2$O capsule is not removed before each laser shot, and the distance between the Al target and the D$_2$O capsule is 0.5~mm or 1.2~mm for L4694 or L4701, respectively. The exposure time is 40~min for the both shots. Although the capsule expands the quasi-monoenergetic peaks and make low-energy backgrounds, their energy spectra show clearly the peaks at 11.2~MeV with an energy spread of 10.7\% and 12.8 MeV with an energy spread of 13.2\%. These results suggest the possibility that this method may have the high repeatability of the deuteron energy spectra. 

\begin{figure}
\includegraphics[width=7.5cm]{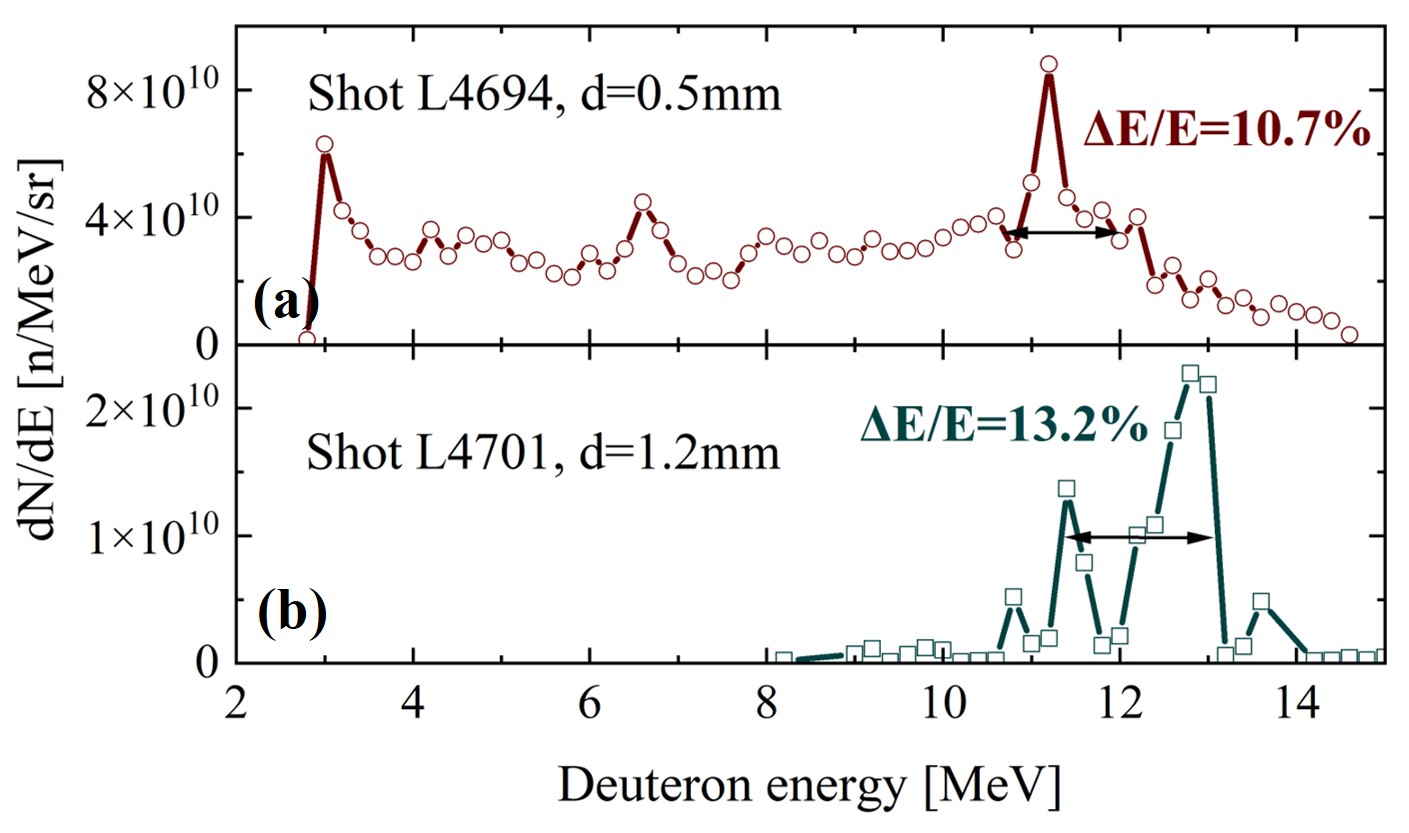}
\caption{
(a) Deuteron spectra for shot L4694, peak deuterons are measured at 11.2~MeV with an energy spread of 10.7\%. (b) Deuteron spectra for shot L4702, peak deuterons are measured at 12.8~MeV with an energy spread of 13.2\%. }
\label{fAPPEDIX}
\end{figure}

\section{Discussion}

The energy spreads  
and conversion efficiencies (CE)
of ions accelerated by laser-plasma interactions with relativistic laser (>10$^{18}$~W/cm$^2$) in the recent experiments are summarized in the Table~\ref{t}.
In the present experiment, we obtained 10.8~MeV deuterons with an energy spread of 4.6$\%$,
which is narrow compered to the previous values in Table ~\ref{t}.

\begin{table*}
\caption{The energy spreads 
and conversion efficiencies
of ions accelerated by laser-plasma interactions with relativistic laser (>10$^{18}~$W/cm$^2$) in experiments.}
\label{t}
\begin{ruledtabular}
\begin{tabular}{ccccccccccccccc}
 Reference &  Pulse duration, Intensity &  Target & Ion Specie & $\Delta $E/E & CE\\
\hline
      A. Maksimchuk et al.\cite{maksimchuk2004high}  &  $\sim$100~fs, $5\times10^{18}$~W/cm$^2$ & 12.5~$\mu$m Al film& H$^{+}$ & 24\% & -\\
     H. Schwoerer et al.\cite{schwoerer2006laser}  &  80~fs, $3\times10^{19}$~W/cm$^2$ & microstructured target& H$^{+}$ & 25\%& -\\
       B. M. Hegelich et al.\cite{hegelich2006laser}  &  0.6~ps, $1\times10^{19}$~W/cm$^2$ & $\sim$1100K Pd foil& C$^{5+}$ & 17\%& -\\
       S. Ter-Avetisyan et al.\cite{ter2006quasimonoenergetic}& 40~fs, $1\times10^{19}$~W/cm$^2$ & 10~$\mu$m D$_2$O droplet & D$^+$ & 14\%& -\\
        A. Henig et al.\cite{henig2009radiation}& 45~fs, $5\times10^{19}$~W/cm$^2$ & 5.3~nm diamondlike carbon&  C$^{6+}$ & 50\%& 2.5\%\\
        
       B. Ramakrishna et al.\cite{ramakrishna2010laser} & 45~fs, $5\times10^{19}$~W/cm$^2$ & 150~nm H$_2$O droplet & H$^+$ & 10\%& -\\
        D. Jung et al.\cite{jung2011monoenergetic} & 0.5~ps, $2\times10^{20}$~W/cm$^2$ &  5nm diamondlike carbon & C$^{6+}$ & 35\%& 0.06\%\\
        
        S. Kar et al.\cite{kar2012ion} & 0.8~ps, $1.25\times10^{20}$~W/cm$^2$ &  50~nm Cu foil & C$^{6+}$ & 60\%& 1\%\\
        
        S. Palaniyappan et al.\cite{palaniyappan2015efficient} & 0.65~ps, $2\times10^{20}$~W/cm$^2$ &  110~nm Al foil & Al$^{11+}$ & 7\% & 5\%\\
        A. Pak et al.\cite{pak2018collisionless} & 1~ps, $1\times10^{20}$~W/cm$^2$ &  ablated Mylar foil & C$^{6+}\&$O$^{8+}$ &10\%& -\\
        G. G. Scott et al.\cite{scott2018dual} & 1.2~ps, $1\times10^{20}$~W/cm$^2$ &    cryogenic target  & D$^+$ & 57\%& 0.5\%\\
         P. Hilz et al.\cite{hilz2018isolated} & 0.5~ps, $7\times10^{20}$~W/cm$^2$ &    plastic sphere target  & H$^+$ & 25\%& 0.3\%\\
         J. Li et al.\cite{li2019laser}& 0.6~ps, $6\times10^{20}$~W/cm$^2$  &  100~nm Ti foil &  Ti$^{20+}$ & 14\%& -\\
         S. Bagchi et al.\cite{bagchi2021quasi}& 25~fs, $1.5\times10^{19}$~W/cm$^2$ &  multi-layer target & Au ion & 20\%& -\\
         H. Ahmed et al.\cite{ahmed2021high}& 1~ps, $3.5\times10^{20}$~W/cm$^2$ &  coil-target & H$^{+}$ & 10\%& -\\
         This work& 1.5~ps, $1\times10^{19}$~W/cm$^2$ &  Al-D$_2$O & D$^+$ & 4.6\%& 0.04\%\\
\end{tabular}
\end{ruledtabular}
\end{table*}

\begin{figure*}[!th]

\includegraphics[width=15cm]{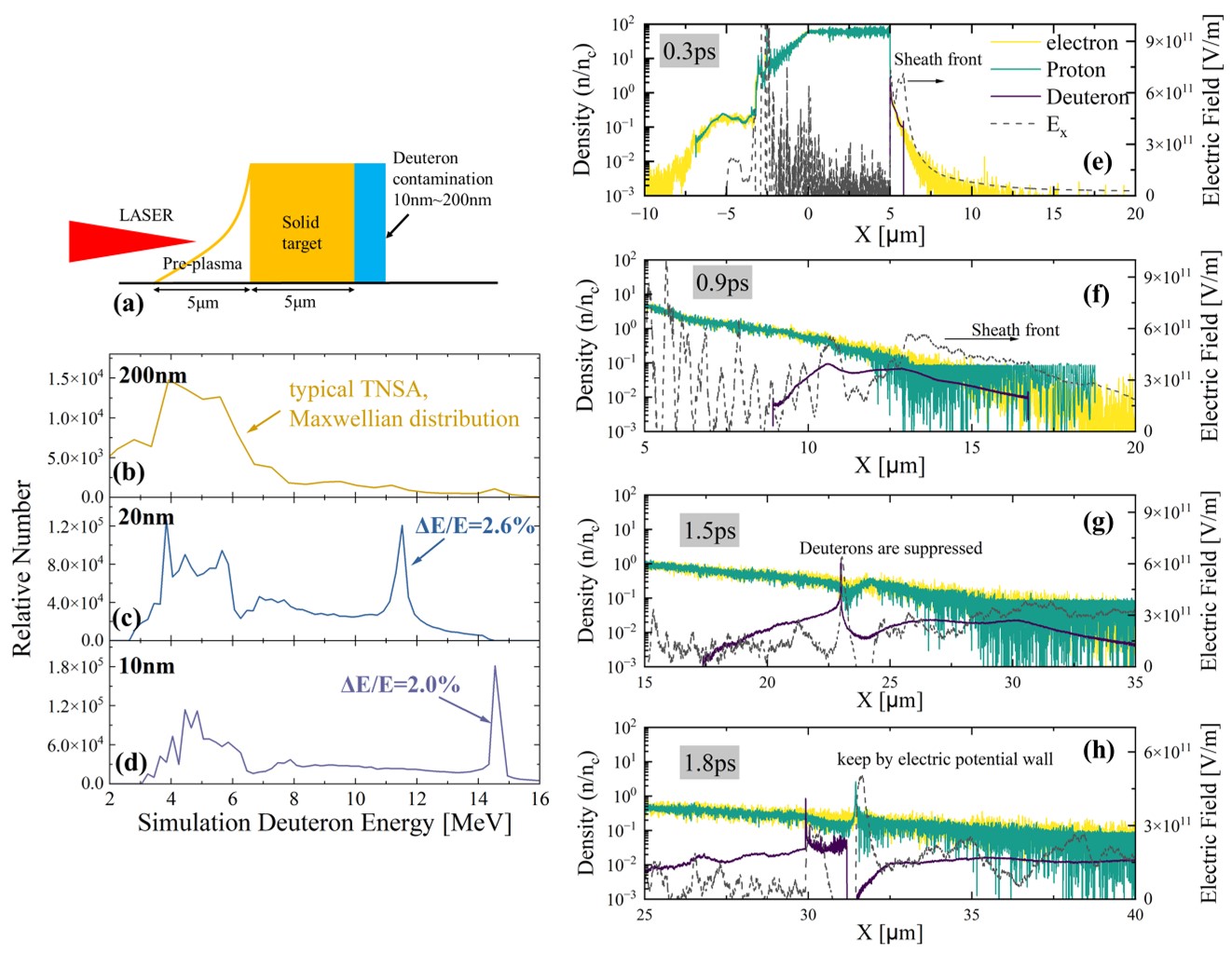}
\caption{
(a) Setup for PIC simulation. Laser irradiates on the solid target consisting of hydrogen with the contamination layer of D$_2$O at the rear side.
(b) Calculated deuteron energy spectrum at 1.5~ps for the D$_2$O-layer thickness of 200~nm. (c) Calculated deuteron energy spectrum at 1.5~ps for the D$_2$O-layer thickness of 20~nm. The spectrum shows the peak with a spread of 2.6\% at 11.5 MeV. (d) Calculated deuteron energy spectrum at 1.5~ps for the D$_2$O-layer thickness of 10~nm. The spectrum shows the peak with a spread of 2.0\% at 14.5 MeV.
(e) Density profile 
for 20~nm thickness
at 0.3~ps. 
The main laser pulse incident into the pre-plasma near -3~$\mu$m.
The sheath field (black dash) is formed by hot electrons (yellow, light gray), deuterons (purple, dark gray) are first accelerated. 
(f) Density profile 
for 20~nm thickness
at 0.9~ps. The deuterons are detached from the solid target and the sheath front moves to protons (green, middle gray). 
(g) Density profile 
for 20~nm thickness
at 1.5~ps. The deuterons are suppressed by the protons and the narrow energy peak is formed.
(h) Density profile for 20~nm thickness  at 1.8~ps. The narrow energy peak of the deuterons is kept by the electric potential wall formed by the faster protons. "
}
\label{fSIM}

\end{figure*}

To investigate the acceleration mechanism beyond a few ps, we conduct 
PIC simulations 
using the EPOCH code\cite{bennett2017users}.
It is reported\cite{iwata2021lateral} that the electron heating and the ion acceleration are degenerated into quasi-1D situation at the normal direction when $w/L\gg1$, where $w$ is the laser focus spot size and $L$ is the thickness of a solid target. Under the present condition of $w~=~50~\mu$m and $L~=~5~\mu$m, the 1D simulations are utilized to interpret our experiment results.
The simulation setup is shown in Fig.~\ref{fSIM}(a). 
A solid target containing protons with a maximum density of 60~n$_c$ is utilized, 
where n$_c$ is the plasma critical density of n$_c=1\times10^{21}$~cm$^{-3}$.
The simulation calculations are conducted without Al ions, because the Al ions are accelerated in late frame from the rear side and they do not affect significantly to the deuteron acceleration.
A contamination layer containing deuterons with a density of 15~n$_c$ is positioned at the rear side of the target.
The thickness of this layer is assumed to be 200~nm, 20~nm, or 10~nm. 
The solid target is accompanied by a pre-plasma.
The laser intensity is set at $1\times10^{19}$~W/cm$^2$ and a pulse duration is 1.5~ps as typical parameters of the LFEX laser. 
The simulation calculation is employed using a mesh size of $\delta x$ = 2~nm 
for 20~nm and 10~nm thicknesses 
to ensure adequate spatial resolution.

The calculated deuteron spectra at 1.5~ps for the three initial thicknesses are shown in Figs. 4(b)-(d). In the case of the 200~nm thickness, the deuteron energy spectrum shows a typical TNSA Maxwellian distribution, whereas quasi-monoenergetic peaks appear for the 20~nm and 10~nm thicknesses. The peak energies are 11.5~MeV with an energy spread of 2.6\% and 14.5~MeV with 2.0\% spread for the 20~nm and 10~nm thicknesses, respectively. These results can explain the experimental results. The peak energy at the thickness of 20~nm is closer to the spread of the measured peak in the L5163 shot than the other calculated results. This result indicates that the thicknesses of the D$_2$O layers are in order of tens of nanometers.
The calculated results show that the deuterons exist at low energy region of 4-6 MeV, but the experimental results of the L5163 and L4701 shots show very low components in this energy region. This discrepancy is an unresolved question, but it suggests that when narrow peaks of deuterons are formed the low energy deuterons are effectively accelerated or decelerated.

The four snapshots
of the density profiles 
for the 20~nm thickness 
show the process of the high energy peak formation. 
At early time of 0.3~ps [Fig.\ref{fSIM} (c)],
the hot electrons form the sheath electric field at the rear side of the target. 
The field strength is approximately $8\times10^{11}$~V/m, which is in agreement with typical field strengths of the TNSA. 
The deuterons in the contamination layer at the rear side are first accelerated by the sheath field at this frame. 
Next, all deuterons in the ultra-thin deuteron layer are accelerated and detached from the solid target at 0.9~ps [Fig.~\ref{fSIM}(d)].
The protons in the solid target are simultaneously accelerated to velocities higher than those of the deuterons because
the mass of the proton is lower than that of the deuteron.
As a result, deuteron acceleration is suppressed by Coulomb interaction with the higher speed protons at 1.5~ps [Fig.\ref{fSIM} (e)], resulting in the quasi-monoenergetic deuteron pulse. 
In the later time, the quasi-monoenergetic energy of the deuterons is kept by the electric potential formed by the faster protons [see Fig.~\ref{fSIM} (h)]. 
From the simulation results, we conclude that deuterons can be accelerated into a quasi-monoenergetic pulse by the following three reasons. First, the deuteron layer is thin enough, so that almost all deuterons can be accelerated from the solid target.
Second, the laser pulse is relatively long ($\sim$ps), making the sheath field for a relatively long time. Third, deuterons deuteron covering on the protons are accelerated in the former frame.

The laser-to-deuteron conversion efficiency in the present experiment [shot L5163] is estimated to 0.04\% by assuming the divergence angle as 1~sr. This value is lower than the reported values in Table \ref{t}.
Because protons have an important role for the generation of narrow deuteron pulses, it is difficult to increase the fraction of deuterons in all accelerated ions.
However, this method is suitable for applications requiring narrow peak pulses rather than high flux,
and this method can be applied to accelerate other ion species into quasi-mono energies.
One of the candidates is quasi-monoenergetic triton acceleration. Recently, continuous energy triton acceleration was demonstrated using the OMEGA EP laser for the study of the nuclear physics of $^6$He through T+T nuclear reactions and nuclear fusion through T+D reactions\cite{schwemmlein2022first}. Generation of quasi-monoenergetic tritons is desirable for further studies of these topics.
Because this method requires the accumulation time in order of several tens of minutes for generation of a D$_2$O layer, this is suitable for low-frequency laser, and other methods such as thin metal foils covered with ultra-thin deuterium-containing layers\cite{prencipe2017targets}  are more suitable for high-frequency laser. 

   \section{Conclusion}
In conclusion, we have reported a method to accelerate quasi-monoenergetic ions via the TNSA mechanism. 
By depositing a thin D$_2$O layer with a thickness of several tens of nm on a metal target as a contamination layer, deuterons are accelerated first by the sheath field, and then compressed by the accelerated protons. As a result, a quasi-monoenergic deuteron pulse is generated.
We obtain quasi-monoenergetic deuteron pulse with an energy of 10.8~MeV and a spread of $\Delta $E/E = 4.6\% using the TPIS in the best shot.
Furthermore, it is expected that quasi-monoenergetic pulses of other type of ions can be generated using a suitable liquid capsule such as super-heavy water (T$_2$O) for triton acceleration.

\nocite{*}

\section*{Data Availability Statement}

The data that support the findings of this study are available from the corresponding author upon reasonable request.

\begin{acknowledgments}
The authors thank the technical support staff of ILE for their assistance with the laser operation, target fabrication and plasma diagnostics. This work was supported by the Collaboration Research Program of ILE, Osaka University. 
This work is funded by JSPS KAKENHI Grant-in-Aid for Scientific Research (JP25420911, JP26246043, JP22H02007, JP22H01239), JST A-STEP (AS2721002c) and JST PRESTO (JPMJPR15PD).T. Wei is surpported by JST SPRING (JPMJSP2138).
\end{acknowledgments}

\nocite{*}
\bibliography{aipsamp}

\begin{thebibliography}{45}%
\makeatletter
\providecommand \@ifxundefined [1]{%
 \@ifx{#1\undefined}
}%
\providecommand \@ifnum [1]{%
 \ifnum #1\expandafter \@firstoftwo
 \else \expandafter \@secondoftwo
 \fi
}%
\providecommand \@ifx [1]{%
 \ifx #1\expandafter \@firstoftwo
 \else \expandafter \@secondoftwo
 \fi
}%
\providecommand \natexlab [1]{#1}%
\providecommand \enquote  [1]{``#1''}%
\providecommand \bibnamefont  [1]{#1}%
\providecommand \bibfnamefont [1]{#1}%
\providecommand \citenamefont [1]{#1}%
\providecommand \href@noop [0]{\@secondoftwo}%
\providecommand \href [0]{\begingroup \@sanitize@url \@href}%
\providecommand \@href[1]{\@@startlink{#1}\@@href}%
\providecommand \@@href[1]{\endgroup#1\@@endlink}%
\providecommand \@sanitize@url [0]{\catcode `\\12\catcode `\$12\catcode
  `\&12\catcode `\#12\catcode `\^12\catcode `\_12\catcode `\%12\relax}%
\providecommand \@@startlink[1]{}%
\providecommand \@@endlink[0]{}%
\providecommand \url  [0]{\begingroup\@sanitize@url \@url }%
\providecommand \@url [1]{\endgroup\@href {#1}{\urlprefix }}%
\providecommand \urlprefix  [0]{URL }%
\providecommand \Eprint [0]{\href }%
\providecommand \doibase [0]{http://dx.doi.org/}%
\providecommand \selectlanguage [0]{\@gobble}%
\providecommand \bibinfo  [0]{\@secondoftwo}%
\providecommand \bibfield  [0]{\@secondoftwo}%
\providecommand \translation [1]{[#1]}%
\providecommand \BibitemOpen [0]{}%
\providecommand \bibitemStop [0]{}%
\providecommand \bibitemNoStop [0]{.\EOS\space}%
\providecommand \EOS [0]{\spacefactor3000\relax}%
\providecommand \BibitemShut  [1]{\csname bibitem#1\endcsname}%
\let\auto@bib@innerbib\@empty
\bibitem [{\citenamefont {Danson}\ \emph {et~al.}(2019)\citenamefont {Danson},
  \citenamefont {Haefner}, \citenamefont {Bromage}, \citenamefont {Butcher},
  \citenamefont {Chanteloup}, \citenamefont {Chowdhury}, \citenamefont
  {Galvanauskas}, \citenamefont {Gizzi}, \citenamefont {Hein}, \citenamefont
  {Hillier} \emph {et~al.}}]{danson2019petawatt}%
  \BibitemOpen
  \bibfield  {author} {\bibinfo {author} {\bibfnamefont {C.~N.}\ \bibnamefont
  {Danson}}, \bibinfo {author} {\bibfnamefont {C.}~\bibnamefont {Haefner}},
  \bibinfo {author} {\bibfnamefont {J.}~\bibnamefont {Bromage}}, \bibinfo
  {author} {\bibfnamefont {T.}~\bibnamefont {Butcher}}, \bibinfo {author}
  {\bibfnamefont {J.-C.~F.}\ \bibnamefont {Chanteloup}}, \bibinfo {author}
  {\bibfnamefont {E.~A.}\ \bibnamefont {Chowdhury}}, \bibinfo {author}
  {\bibfnamefont {A.}~\bibnamefont {Galvanauskas}}, \bibinfo {author}
  {\bibfnamefont {L.~A.}\ \bibnamefont {Gizzi}}, \bibinfo {author}
  {\bibfnamefont {J.}~\bibnamefont {Hein}}, \bibinfo {author} {\bibfnamefont
  {D.~I.}\ \bibnamefont {Hillier}},  \emph {et~al.},\ }\bibfield  {title}
  {\enquote {\bibinfo {title} {Petawatt and exawatt class lasers worldwide},}\
  }\href@noop {} {\bibfield  {journal} {\bibinfo  {journal} {High Power Laser
  Science and Engineering}\ }\textbf {\bibinfo {volume} {7}},\ \bibinfo {pages}
  {e54} (\bibinfo {year} {2019})}\BibitemShut {NoStop}%
\bibitem [{\citenamefont {Macchi}, \citenamefont {Borghesi},\ and\
  \citenamefont {Passoni}(2013)}]{macchi2013ion}%
  \BibitemOpen
  \bibfield  {author} {\bibinfo {author} {\bibfnamefont {A.}~\bibnamefont
  {Macchi}}, \bibinfo {author} {\bibfnamefont {M.}~\bibnamefont {Borghesi}}, \
  and\ \bibinfo {author} {\bibfnamefont {M.}~\bibnamefont {Passoni}},\
  }\bibfield  {title} {\enquote {\bibinfo {title} {Ion acceleration by
  superintense laser-plasma interaction},}\ }\href@noop {} {\bibfield
  {journal} {\bibinfo  {journal} {Reviews of Modern Physics}\ }\textbf
  {\bibinfo {volume} {85}},\ \bibinfo {pages} {751} (\bibinfo {year}
  {2013})}\BibitemShut {NoStop}%
\bibitem [{\citenamefont {Higginson}\ \emph {et~al.}(2018)\citenamefont
  {Higginson}, \citenamefont {Gray}, \citenamefont {King}, \citenamefont
  {Dance}, \citenamefont {Williamson}, \citenamefont {Butler}, \citenamefont
  {Wilson}, \citenamefont {Capdessus}, \citenamefont {Armstrong}, \citenamefont
  {Green} \emph {et~al.}}]{higginson2018near}%
  \BibitemOpen
  \bibfield  {author} {\bibinfo {author} {\bibfnamefont {A.}~\bibnamefont
  {Higginson}}, \bibinfo {author} {\bibfnamefont {R.}~\bibnamefont {Gray}},
  \bibinfo {author} {\bibfnamefont {M.}~\bibnamefont {King}}, \bibinfo {author}
  {\bibfnamefont {R.}~\bibnamefont {Dance}}, \bibinfo {author} {\bibfnamefont
  {S.}~\bibnamefont {Williamson}}, \bibinfo {author} {\bibfnamefont
  {N.}~\bibnamefont {Butler}}, \bibinfo {author} {\bibfnamefont
  {R.}~\bibnamefont {Wilson}}, \bibinfo {author} {\bibfnamefont
  {R.}~\bibnamefont {Capdessus}}, \bibinfo {author} {\bibfnamefont
  {C.}~\bibnamefont {Armstrong}}, \bibinfo {author} {\bibfnamefont
  {J.}~\bibnamefont {Green}},  \emph {et~al.},\ }\bibfield  {title} {\enquote
  {\bibinfo {title} {{Near-100 MeV protons via a laser-driven
  transparency-enhanced hybrid acceleration scheme}},}\ }\href@noop {}
  {\bibfield  {journal} {\bibinfo  {journal} {Nature communications}\ }\textbf
  {\bibinfo {volume} {9}},\ \bibinfo {pages} {1--9} (\bibinfo {year}
  {2018})}\BibitemShut {NoStop}%
\bibitem [{\citenamefont {Yogo}\ \emph {et~al.}(2017)\citenamefont {Yogo},
  \citenamefont {Mima}, \citenamefont {Iwata}, \citenamefont {Tosaki},
  \citenamefont {Morace}, \citenamefont {Arikawa}, \citenamefont {Fujioka},
  \citenamefont {Johzaki}, \citenamefont {Sentoku}, \citenamefont {Nishimura}
  \emph {et~al.}}]{yogo2017boosting}%
  \BibitemOpen
  \bibfield  {author} {\bibinfo {author} {\bibfnamefont {A.}~\bibnamefont
  {Yogo}}, \bibinfo {author} {\bibfnamefont {K.}~\bibnamefont {Mima}}, \bibinfo
  {author} {\bibfnamefont {N.}~\bibnamefont {Iwata}}, \bibinfo {author}
  {\bibfnamefont {S.}~\bibnamefont {Tosaki}}, \bibinfo {author} {\bibfnamefont
  {A.}~\bibnamefont {Morace}}, \bibinfo {author} {\bibfnamefont
  {Y.}~\bibnamefont {Arikawa}}, \bibinfo {author} {\bibfnamefont
  {S.}~\bibnamefont {Fujioka}}, \bibinfo {author} {\bibfnamefont
  {T.}~\bibnamefont {Johzaki}}, \bibinfo {author} {\bibfnamefont
  {Y.}~\bibnamefont {Sentoku}}, \bibinfo {author} {\bibfnamefont
  {H.}~\bibnamefont {Nishimura}},  \emph {et~al.},\ }\bibfield  {title}
  {\enquote {\bibinfo {title} {Boosting laser-ion acceleration with
  multi-picosecond pulses},}\ }\href@noop {} {\bibfield  {journal} {\bibinfo
  {journal} {Scientific reports}\ }\textbf {\bibinfo {volume} {7}},\ \bibinfo
  {pages} {1--10} (\bibinfo {year} {2017})}\BibitemShut {NoStop}%
\bibitem [{\citenamefont {Yogo}\ \emph {et~al.}(2023)\citenamefont {Yogo},
  \citenamefont {Lan}, \citenamefont {Arikawa}, \citenamefont {Abe},
  \citenamefont {Mirfayzi}, \citenamefont {Wei}, \citenamefont {Mori},
  \citenamefont {Golovin}, \citenamefont {Hayakawa}, \citenamefont {Iwata}
  \emph {et~al.}}]{yogo2023laser}%
  \BibitemOpen
  \bibfield  {author} {\bibinfo {author} {\bibfnamefont {A.}~\bibnamefont
  {Yogo}}, \bibinfo {author} {\bibfnamefont {Z.}~\bibnamefont {Lan}}, \bibinfo
  {author} {\bibfnamefont {Y.}~\bibnamefont {Arikawa}}, \bibinfo {author}
  {\bibfnamefont {Y.}~\bibnamefont {Abe}}, \bibinfo {author} {\bibfnamefont
  {S.}~\bibnamefont {Mirfayzi}}, \bibinfo {author} {\bibfnamefont
  {T.}~\bibnamefont {Wei}}, \bibinfo {author} {\bibfnamefont {T.}~\bibnamefont
  {Mori}}, \bibinfo {author} {\bibfnamefont {D.}~\bibnamefont {Golovin}},
  \bibinfo {author} {\bibfnamefont {T.}~\bibnamefont {Hayakawa}}, \bibinfo
  {author} {\bibfnamefont {N.}~\bibnamefont {Iwata}},  \emph {et~al.},\
  }\bibfield  {title} {\enquote {\bibinfo {title} {Laser-driven neutron
  generation realizing single-shot resonance spectroscopy},}\ }\href@noop {}
  {\bibfield  {journal} {\bibinfo  {journal} {Physical Review X}\ }\textbf
  {\bibinfo {volume} {13}},\ \bibinfo {pages} {011011} (\bibinfo {year}
  {2023})}\BibitemShut {NoStop}%
\bibitem [{\citenamefont {Romagnani}\ \emph {et~al.}(2008)\citenamefont
  {Romagnani}, \citenamefont {Borghesi}, \citenamefont {Cecchetti},
  \citenamefont {Kar}, \citenamefont {Antici}, \citenamefont {Audebert},
  \citenamefont {Bandhoupadjay}, \citenamefont {Ceccherini}, \citenamefont
  {Cowan}, \citenamefont {Fuchs} \emph {et~al.}}]{romagnani2008proton}%
  \BibitemOpen
  \bibfield  {author} {\bibinfo {author} {\bibfnamefont {L.}~\bibnamefont
  {Romagnani}}, \bibinfo {author} {\bibfnamefont {M.}~\bibnamefont {Borghesi}},
  \bibinfo {author} {\bibfnamefont {C.}~\bibnamefont {Cecchetti}}, \bibinfo
  {author} {\bibfnamefont {S.}~\bibnamefont {Kar}}, \bibinfo {author}
  {\bibfnamefont {P.}~\bibnamefont {Antici}}, \bibinfo {author} {\bibfnamefont
  {P.}~\bibnamefont {Audebert}}, \bibinfo {author} {\bibfnamefont
  {S.}~\bibnamefont {Bandhoupadjay}}, \bibinfo {author} {\bibfnamefont
  {F.}~\bibnamefont {Ceccherini}}, \bibinfo {author} {\bibfnamefont
  {T.}~\bibnamefont {Cowan}}, \bibinfo {author} {\bibfnamefont
  {J.}~\bibnamefont {Fuchs}},  \emph {et~al.},\ }\bibfield  {title} {\enquote
  {\bibinfo {title} {Proton probing measurement of electric and magnetic fields
  generated by ns and ps laser-matter interactions},}\ }\href@noop {}
  {\bibfield  {journal} {\bibinfo  {journal} {Laser and Particle Beams}\
  }\textbf {\bibinfo {volume} {26}},\ \bibinfo {pages} {241--248} (\bibinfo
  {year} {2008})}\BibitemShut {NoStop}%
\bibitem [{\citenamefont {Bulanov}\ and\ \citenamefont
  {Khoroshkov}(2002)}]{bulanov2002feasibility}%
  \BibitemOpen
  \bibfield  {author} {\bibinfo {author} {\bibfnamefont {S.}~\bibnamefont
  {Bulanov}}\ and\ \bibinfo {author} {\bibfnamefont {V.}~\bibnamefont
  {Khoroshkov}},\ }\bibfield  {title} {\enquote {\bibinfo {title} {Feasibility
  of using laser ion accelerators in proton therapy},}\ }\href@noop {}
  {\bibfield  {journal} {\bibinfo  {journal} {Plasma Physics Reports}\ }\textbf
  {\bibinfo {volume} {28}},\ \bibinfo {pages} {453--456} (\bibinfo {year}
  {2002})}\BibitemShut {NoStop}%
\bibitem [{\citenamefont {Ledingham}\ \emph {et~al.}(2014)\citenamefont
  {Ledingham}, \citenamefont {Bolton}, \citenamefont {Shikazono},\ and\
  \citenamefont {Ma}}]{ledingham2014towards}%
  \BibitemOpen
  \bibfield  {author} {\bibinfo {author} {\bibfnamefont {K.~W.}\ \bibnamefont
  {Ledingham}}, \bibinfo {author} {\bibfnamefont {P.~R.}\ \bibnamefont
  {Bolton}}, \bibinfo {author} {\bibfnamefont {N.}~\bibnamefont {Shikazono}}, \
  and\ \bibinfo {author} {\bibfnamefont {C.-M.~C.}\ \bibnamefont {Ma}},\
  }\bibfield  {title} {\enquote {\bibinfo {title} {Towards laser driven hadron
  cancer radiotherapy: A review of progress},}\ }\href@noop {} {\bibfield
  {journal} {\bibinfo  {journal} {Applied Sciences}\ }\textbf {\bibinfo
  {volume} {4}},\ \bibinfo {pages} {402--443} (\bibinfo {year}
  {2014})}\BibitemShut {NoStop}%
\bibitem [{\citenamefont {Bulanov}\ \emph {et~al.}(2014)\citenamefont
  {Bulanov}, \citenamefont {Wilkens}, \citenamefont {Esirkepov}, \citenamefont
  {Korn}, \citenamefont {Kraft}, \citenamefont {Kraft}, \citenamefont {Molls},\
  and\ \citenamefont {Khoroshkov}}]{bulanov2014laser}%
  \BibitemOpen
  \bibfield  {author} {\bibinfo {author} {\bibfnamefont {S.~V.}\ \bibnamefont
  {Bulanov}}, \bibinfo {author} {\bibfnamefont {J.~J.}\ \bibnamefont
  {Wilkens}}, \bibinfo {author} {\bibfnamefont {T.~Z.}\ \bibnamefont
  {Esirkepov}}, \bibinfo {author} {\bibfnamefont {G.}~\bibnamefont {Korn}},
  \bibinfo {author} {\bibfnamefont {G.}~\bibnamefont {Kraft}}, \bibinfo
  {author} {\bibfnamefont {S.~D.}\ \bibnamefont {Kraft}}, \bibinfo {author}
  {\bibfnamefont {M.}~\bibnamefont {Molls}}, \ and\ \bibinfo {author}
  {\bibfnamefont {V.~S.}\ \bibnamefont {Khoroshkov}},\ }\bibfield  {title}
  {\enquote {\bibinfo {title} {Laser ion acceleration for hadron therapy},}\
  }\href@noop {} {\bibfield  {journal} {\bibinfo  {journal} {Physics-Uspekhi}\
  }\textbf {\bibinfo {volume} {57}},\ \bibinfo {pages} {1149} (\bibinfo {year}
  {2014})}\BibitemShut {NoStop}%
\bibitem [{\citenamefont {Roth}\ \emph {et~al.}(2001)\citenamefont {Roth},
  \citenamefont {Cowan}, \citenamefont {Key}, \citenamefont {Hatchett},
  \citenamefont {Brown}, \citenamefont {Fountain}, \citenamefont {Johnson},
  \citenamefont {Pennington}, \citenamefont {Snavely}, \citenamefont {Wilks}
  \emph {et~al.}}]{roth2001fast}%
  \BibitemOpen
  \bibfield  {author} {\bibinfo {author} {\bibfnamefont {M.}~\bibnamefont
  {Roth}}, \bibinfo {author} {\bibfnamefont {T.}~\bibnamefont {Cowan}},
  \bibinfo {author} {\bibfnamefont {M.}~\bibnamefont {Key}}, \bibinfo {author}
  {\bibfnamefont {S.}~\bibnamefont {Hatchett}}, \bibinfo {author}
  {\bibfnamefont {C.}~\bibnamefont {Brown}}, \bibinfo {author} {\bibfnamefont
  {W.}~\bibnamefont {Fountain}}, \bibinfo {author} {\bibfnamefont
  {J.}~\bibnamefont {Johnson}}, \bibinfo {author} {\bibfnamefont
  {D.}~\bibnamefont {Pennington}}, \bibinfo {author} {\bibfnamefont
  {R.}~\bibnamefont {Snavely}}, \bibinfo {author} {\bibfnamefont
  {S.}~\bibnamefont {Wilks}},  \emph {et~al.},\ }\bibfield  {title} {\enquote
  {\bibinfo {title} {Fast ignition by intense laser-accelerated proton
  beams},}\ }\href@noop {} {\bibfield  {journal} {\bibinfo  {journal} {Physical
  review letters}\ }\textbf {\bibinfo {volume} {86}},\ \bibinfo {pages} {436}
  (\bibinfo {year} {2001})}\BibitemShut {NoStop}%
\bibitem [{\citenamefont {Fern{\'a}ndez}\ \emph {et~al.}(2009)\citenamefont
  {Fern{\'a}ndez}, \citenamefont {Honrubia}, \citenamefont {Albright},
  \citenamefont {Flippo}, \citenamefont {Gautier}, \citenamefont {Hegelich},
  \citenamefont {Schmitt}, \citenamefont {Temporal},\ and\ \citenamefont
  {Yin}}]{fernandez2009progress}%
  \BibitemOpen
  \bibfield  {author} {\bibinfo {author} {\bibfnamefont {J.~C.}\ \bibnamefont
  {Fern{\'a}ndez}}, \bibinfo {author} {\bibfnamefont {J.}~\bibnamefont
  {Honrubia}}, \bibinfo {author} {\bibfnamefont {B.~J.}\ \bibnamefont
  {Albright}}, \bibinfo {author} {\bibfnamefont {K.~A.}\ \bibnamefont
  {Flippo}}, \bibinfo {author} {\bibfnamefont {D.~C.}\ \bibnamefont {Gautier}},
  \bibinfo {author} {\bibfnamefont {B.~M.}\ \bibnamefont {Hegelich}}, \bibinfo
  {author} {\bibfnamefont {M.~J.}\ \bibnamefont {Schmitt}}, \bibinfo {author}
  {\bibfnamefont {M.}~\bibnamefont {Temporal}}, \ and\ \bibinfo {author}
  {\bibfnamefont {L.}~\bibnamefont {Yin}},\ }\bibfield  {title} {\enquote
  {\bibinfo {title} {Progress and prospects of ion-driven fast ignition},}\
  }\href@noop {} {\bibfield  {journal} {\bibinfo  {journal} {Nuclear fusion}\
  }\textbf {\bibinfo {volume} {49}},\ \bibinfo {pages} {065004} (\bibinfo
  {year} {2009})}\BibitemShut {NoStop}%
\bibitem [{\citenamefont {Alejo}\ \emph {et~al.}(2017)\citenamefont {Alejo},
  \citenamefont {Krygier}, \citenamefont {Ahmed}, \citenamefont {Morrison},
  \citenamefont {Clarke}, \citenamefont {Fuchs}, \citenamefont {Green},
  \citenamefont {Green}, \citenamefont {Jung}, \citenamefont {Kleinschmidt}
  \emph {et~al.}}]{alejo2017high}%
  \BibitemOpen
  \bibfield  {author} {\bibinfo {author} {\bibfnamefont {A.}~\bibnamefont
  {Alejo}}, \bibinfo {author} {\bibfnamefont {A.}~\bibnamefont {Krygier}},
  \bibinfo {author} {\bibfnamefont {H.}~\bibnamefont {Ahmed}}, \bibinfo
  {author} {\bibfnamefont {J.}~\bibnamefont {Morrison}}, \bibinfo {author}
  {\bibfnamefont {R.}~\bibnamefont {Clarke}}, \bibinfo {author} {\bibfnamefont
  {J.}~\bibnamefont {Fuchs}}, \bibinfo {author} {\bibfnamefont
  {A.}~\bibnamefont {Green}}, \bibinfo {author} {\bibfnamefont
  {J.}~\bibnamefont {Green}}, \bibinfo {author} {\bibfnamefont
  {D.}~\bibnamefont {Jung}}, \bibinfo {author} {\bibfnamefont {A.}~\bibnamefont
  {Kleinschmidt}},  \emph {et~al.},\ }\bibfield  {title} {\enquote {\bibinfo
  {title} {High flux, beamed neutron sources employing deuteron-rich ion beams
  from d2o-ice layered targets},}\ }\href@noop {} {\bibfield  {journal}
  {\bibinfo  {journal} {Plasma Physics and Controlled Fusion}\ }\textbf
  {\bibinfo {volume} {59}},\ \bibinfo {pages} {064004} (\bibinfo {year}
  {2017})}\BibitemShut {NoStop}%
\bibitem [{\citenamefont {Esirkepov}\ \emph {et~al.}(2002)\citenamefont
  {Esirkepov}, \citenamefont {Bulanov}, \citenamefont {Nishihara},
  \citenamefont {Tajima}, \citenamefont {Pegoraro}, \citenamefont {Khoroshkov},
  \citenamefont {Mima}, \citenamefont {Daido}, \citenamefont {Kato},
  \citenamefont {Kitagawa} \emph {et~al.}}]{esirkepov2002proposed}%
  \BibitemOpen
  \bibfield  {author} {\bibinfo {author} {\bibfnamefont {T.~Z.}\ \bibnamefont
  {Esirkepov}}, \bibinfo {author} {\bibfnamefont {S.}~\bibnamefont {Bulanov}},
  \bibinfo {author} {\bibfnamefont {K.}~\bibnamefont {Nishihara}}, \bibinfo
  {author} {\bibfnamefont {T.}~\bibnamefont {Tajima}}, \bibinfo {author}
  {\bibfnamefont {F.}~\bibnamefont {Pegoraro}}, \bibinfo {author}
  {\bibfnamefont {V.}~\bibnamefont {Khoroshkov}}, \bibinfo {author}
  {\bibfnamefont {K.}~\bibnamefont {Mima}}, \bibinfo {author} {\bibfnamefont
  {H.}~\bibnamefont {Daido}}, \bibinfo {author} {\bibfnamefont
  {Y.}~\bibnamefont {Kato}}, \bibinfo {author} {\bibfnamefont {Y.}~\bibnamefont
  {Kitagawa}},  \emph {et~al.},\ }\bibfield  {title} {\enquote {\bibinfo
  {title} {Proposed double-layer target for the generation of high-quality
  laser-accelerated ion beams},}\ }\href@noop {} {\bibfield  {journal}
  {\bibinfo  {journal} {Physical review letters}\ }\textbf {\bibinfo {volume}
  {89}},\ \bibinfo {pages} {175003} (\bibinfo {year} {2002})}\BibitemShut
  {NoStop}%
\bibitem [{\citenamefont {Maksimchuk}\ \emph {et~al.}(2004)\citenamefont
  {Maksimchuk}, \citenamefont {Flippo}, \citenamefont {Krause}, \citenamefont
  {Mourou}, \citenamefont {Nemoto}, \citenamefont {Shultz}, \citenamefont
  {Umstadter}, \citenamefont {Vane}, \citenamefont {Bychenkov}, \citenamefont
  {Dudnikova} \emph {et~al.}}]{maksimchuk2004high}%
  \BibitemOpen
  \bibfield  {author} {\bibinfo {author} {\bibfnamefont {A.}~\bibnamefont
  {Maksimchuk}}, \bibinfo {author} {\bibfnamefont {K.}~\bibnamefont {Flippo}},
  \bibinfo {author} {\bibfnamefont {H.}~\bibnamefont {Krause}}, \bibinfo
  {author} {\bibfnamefont {G.}~\bibnamefont {Mourou}}, \bibinfo {author}
  {\bibfnamefont {K.}~\bibnamefont {Nemoto}}, \bibinfo {author} {\bibfnamefont
  {D.}~\bibnamefont {Shultz}}, \bibinfo {author} {\bibfnamefont
  {D.}~\bibnamefont {Umstadter}}, \bibinfo {author} {\bibfnamefont
  {R.}~\bibnamefont {Vane}}, \bibinfo {author} {\bibfnamefont {V.~Y.}\
  \bibnamefont {Bychenkov}}, \bibinfo {author} {\bibfnamefont {G.}~\bibnamefont
  {Dudnikova}},  \emph {et~al.},\ }\bibfield  {title} {\enquote {\bibinfo
  {title} {High-energy ion generation by short laser pulses},}\ }\href@noop {}
  {\bibfield  {journal} {\bibinfo  {journal} {Plasma Physics Reports}\ }\textbf
  {\bibinfo {volume} {30}},\ \bibinfo {pages} {473--495} (\bibinfo {year}
  {2004})}\BibitemShut {NoStop}%
\bibitem [{\citenamefont {Schwoerer}\ \emph {et~al.}(2006)\citenamefont
  {Schwoerer}, \citenamefont {Pfotenhauer}, \citenamefont {J{\"a}ckel},
  \citenamefont {Amthor}, \citenamefont {Liesfeld}, \citenamefont {Ziegler},
  \citenamefont {Sauerbrey}, \citenamefont {Ledingham},\ and\ \citenamefont
  {Esirkepov}}]{schwoerer2006laser}%
  \BibitemOpen
  \bibfield  {author} {\bibinfo {author} {\bibfnamefont {H.}~\bibnamefont
  {Schwoerer}}, \bibinfo {author} {\bibfnamefont {S.}~\bibnamefont
  {Pfotenhauer}}, \bibinfo {author} {\bibfnamefont {O.}~\bibnamefont
  {J{\"a}ckel}}, \bibinfo {author} {\bibfnamefont {K.-U.}\ \bibnamefont
  {Amthor}}, \bibinfo {author} {\bibfnamefont {B.}~\bibnamefont {Liesfeld}},
  \bibinfo {author} {\bibfnamefont {W.}~\bibnamefont {Ziegler}}, \bibinfo
  {author} {\bibfnamefont {R.}~\bibnamefont {Sauerbrey}}, \bibinfo {author}
  {\bibfnamefont {K.}~\bibnamefont {Ledingham}}, \ and\ \bibinfo {author}
  {\bibfnamefont {T.}~\bibnamefont {Esirkepov}},\ }\bibfield  {title} {\enquote
  {\bibinfo {title} {Laser-plasma acceleration of quasi-monoenergetic protons
  from microstructured targets},}\ }\href@noop {} {\bibfield  {journal}
  {\bibinfo  {journal} {Nature}\ }\textbf {\bibinfo {volume} {439}},\ \bibinfo
  {pages} {445--448} (\bibinfo {year} {2006})}\BibitemShut {NoStop}%
\bibitem [{\citenamefont {Hegelich}\ \emph {et~al.}(2006)\citenamefont
  {Hegelich}, \citenamefont {Albright}, \citenamefont {Cobble}, \citenamefont
  {Flippo}, \citenamefont {Letzring}, \citenamefont {Paffett}, \citenamefont
  {Ruhl}, \citenamefont {Schreiber}, \citenamefont {Schulze},\ and\
  \citenamefont {Fern{\'a}ndez}}]{hegelich2006laser}%
  \BibitemOpen
  \bibfield  {author} {\bibinfo {author} {\bibfnamefont {B.~M.}\ \bibnamefont
  {Hegelich}}, \bibinfo {author} {\bibfnamefont {B.}~\bibnamefont {Albright}},
  \bibinfo {author} {\bibfnamefont {J.}~\bibnamefont {Cobble}}, \bibinfo
  {author} {\bibfnamefont {K.}~\bibnamefont {Flippo}}, \bibinfo {author}
  {\bibfnamefont {S.}~\bibnamefont {Letzring}}, \bibinfo {author}
  {\bibfnamefont {M.}~\bibnamefont {Paffett}}, \bibinfo {author} {\bibfnamefont
  {H.}~\bibnamefont {Ruhl}}, \bibinfo {author} {\bibfnamefont {J.}~\bibnamefont
  {Schreiber}}, \bibinfo {author} {\bibfnamefont {R.}~\bibnamefont {Schulze}},
  \ and\ \bibinfo {author} {\bibfnamefont {J.}~\bibnamefont {Fern{\'a}ndez}},\
  }\bibfield  {title} {\enquote {\bibinfo {title} {{Laser acceleration of
  quasi-monoenergetic MeV ion beams}},}\ }\href@noop {} {\bibfield  {journal}
  {\bibinfo  {journal} {Nature}\ }\textbf {\bibinfo {volume} {439}},\ \bibinfo
  {pages} {441--444} (\bibinfo {year} {2006})}\BibitemShut {NoStop}%
\bibitem [{\citenamefont {Ter-Avetisyan}\ \emph {et~al.}(2006)\citenamefont
  {Ter-Avetisyan}, \citenamefont {Schn{\"u}rer}, \citenamefont {Nickles},
  \citenamefont {Kalashnikov}, \citenamefont {Risse}, \citenamefont {Sokollik},
  \citenamefont {Sandner}, \citenamefont {Andreev},\ and\ \citenamefont
  {Tikhonchuk}}]{ter2006quasimonoenergetic}%
  \BibitemOpen
  \bibfield  {author} {\bibinfo {author} {\bibfnamefont {S.}~\bibnamefont
  {Ter-Avetisyan}}, \bibinfo {author} {\bibfnamefont {M.}~\bibnamefont
  {Schn{\"u}rer}}, \bibinfo {author} {\bibfnamefont {P.}~\bibnamefont
  {Nickles}}, \bibinfo {author} {\bibfnamefont {M.}~\bibnamefont
  {Kalashnikov}}, \bibinfo {author} {\bibfnamefont {E.}~\bibnamefont {Risse}},
  \bibinfo {author} {\bibfnamefont {T.}~\bibnamefont {Sokollik}}, \bibinfo
  {author} {\bibfnamefont {W.}~\bibnamefont {Sandner}}, \bibinfo {author}
  {\bibfnamefont {A.}~\bibnamefont {Andreev}}, \ and\ \bibinfo {author}
  {\bibfnamefont {V.}~\bibnamefont {Tikhonchuk}},\ }\bibfield  {title}
  {\enquote {\bibinfo {title} {Quasimonoenergetic deuteron bursts produced by
  ultraintense laser pulses},}\ }\href@noop {} {\bibfield  {journal} {\bibinfo
  {journal} {Physical Review Letters}\ }\textbf {\bibinfo {volume} {96}},\
  \bibinfo {pages} {145006} (\bibinfo {year} {2006})}\BibitemShut {NoStop}%
\bibitem [{\citenamefont {Henig}\ \emph {et~al.}(2009)\citenamefont {Henig},
  \citenamefont {Steinke}, \citenamefont {Schn{\"u}rer}, \citenamefont
  {Sokollik}, \citenamefont {H{\"o}rlein}, \citenamefont {Kiefer},
  \citenamefont {Jung}, \citenamefont {Schreiber}, \citenamefont {Hegelich},
  \citenamefont {Yan} \emph {et~al.}}]{henig2009radiation}%
  \BibitemOpen
  \bibfield  {author} {\bibinfo {author} {\bibfnamefont {A.}~\bibnamefont
  {Henig}}, \bibinfo {author} {\bibfnamefont {S.}~\bibnamefont {Steinke}},
  \bibinfo {author} {\bibfnamefont {M.}~\bibnamefont {Schn{\"u}rer}}, \bibinfo
  {author} {\bibfnamefont {T.}~\bibnamefont {Sokollik}}, \bibinfo {author}
  {\bibfnamefont {R.}~\bibnamefont {H{\"o}rlein}}, \bibinfo {author}
  {\bibfnamefont {D.}~\bibnamefont {Kiefer}}, \bibinfo {author} {\bibfnamefont
  {D.}~\bibnamefont {Jung}}, \bibinfo {author} {\bibfnamefont {J.}~\bibnamefont
  {Schreiber}}, \bibinfo {author} {\bibfnamefont {B.}~\bibnamefont {Hegelich}},
  \bibinfo {author} {\bibfnamefont {X.}~\bibnamefont {Yan}},  \emph {et~al.},\
  }\bibfield  {title} {\enquote {\bibinfo {title} {Radiation-pressure
  acceleration of ion beams driven by circularly polarized laser pulses},}\
  }\href@noop {} {\bibfield  {journal} {\bibinfo  {journal} {Physical Review
  Letters}\ }\textbf {\bibinfo {volume} {103}},\ \bibinfo {pages} {245003}
  (\bibinfo {year} {2009})}\BibitemShut {NoStop}%
\bibitem [{\citenamefont {Ramakrishna}\ \emph {et~al.}(2010)\citenamefont
  {Ramakrishna}, \citenamefont {Murakami}, \citenamefont {Borghesi},
  \citenamefont {Ehrentraut}, \citenamefont {Nickles}, \citenamefont
  {Schn{\"u}rer}, \citenamefont {Steinke}, \citenamefont {Psikal},
  \citenamefont {Tikhonchuk},\ and\ \citenamefont
  {Ter-Avetisyan}}]{ramakrishna2010laser}%
  \BibitemOpen
  \bibfield  {author} {\bibinfo {author} {\bibfnamefont {B.}~\bibnamefont
  {Ramakrishna}}, \bibinfo {author} {\bibfnamefont {M.}~\bibnamefont
  {Murakami}}, \bibinfo {author} {\bibfnamefont {M.}~\bibnamefont {Borghesi}},
  \bibinfo {author} {\bibfnamefont {L.}~\bibnamefont {Ehrentraut}}, \bibinfo
  {author} {\bibfnamefont {P.}~\bibnamefont {Nickles}}, \bibinfo {author}
  {\bibfnamefont {M.}~\bibnamefont {Schn{\"u}rer}}, \bibinfo {author}
  {\bibfnamefont {S.}~\bibnamefont {Steinke}}, \bibinfo {author} {\bibfnamefont
  {J.}~\bibnamefont {Psikal}}, \bibinfo {author} {\bibfnamefont
  {V.}~\bibnamefont {Tikhonchuk}}, \ and\ \bibinfo {author} {\bibfnamefont
  {S.}~\bibnamefont {Ter-Avetisyan}},\ }\bibfield  {title} {\enquote {\bibinfo
  {title} {Laser-driven quasimonoenergetic proton burst from water spray
  target},}\ }\href@noop {} {\bibfield  {journal} {\bibinfo  {journal} {Physics
  of Plasmas}\ }\textbf {\bibinfo {volume} {17}} (\bibinfo {year}
  {2010})}\BibitemShut {NoStop}%
\bibitem [{\citenamefont {Jung}\ \emph {et~al.}(2011)\citenamefont {Jung},
  \citenamefont {Yin}, \citenamefont {Albright}, \citenamefont {Gautier},
  \citenamefont {H{\"o}rlein}, \citenamefont {Kiefer}, \citenamefont {Henig},
  \citenamefont {Johnson}, \citenamefont {Letzring}, \citenamefont
  {Palaniyappan} \emph {et~al.}}]{jung2011monoenergetic}%
  \BibitemOpen
  \bibfield  {author} {\bibinfo {author} {\bibfnamefont {D.}~\bibnamefont
  {Jung}}, \bibinfo {author} {\bibfnamefont {L.}~\bibnamefont {Yin}}, \bibinfo
  {author} {\bibfnamefont {B.}~\bibnamefont {Albright}}, \bibinfo {author}
  {\bibfnamefont {D.}~\bibnamefont {Gautier}}, \bibinfo {author} {\bibfnamefont
  {R.}~\bibnamefont {H{\"o}rlein}}, \bibinfo {author} {\bibfnamefont
  {D.}~\bibnamefont {Kiefer}}, \bibinfo {author} {\bibfnamefont
  {A.}~\bibnamefont {Henig}}, \bibinfo {author} {\bibfnamefont
  {R.}~\bibnamefont {Johnson}}, \bibinfo {author} {\bibfnamefont
  {S.}~\bibnamefont {Letzring}}, \bibinfo {author} {\bibfnamefont
  {S.}~\bibnamefont {Palaniyappan}},  \emph {et~al.},\ }\bibfield  {title}
  {\enquote {\bibinfo {title} {Monoenergetic ion beam generation by driving ion
  solitary waves with circularly polarized laser light},}\ }\href@noop {}
  {\bibfield  {journal} {\bibinfo  {journal} {Physical Review Letters}\
  }\textbf {\bibinfo {volume} {107}},\ \bibinfo {pages} {115002} (\bibinfo
  {year} {2011})}\BibitemShut {NoStop}%
\bibitem [{\citenamefont {Kar}\ \emph {et~al.}(2012)\citenamefont {Kar},
  \citenamefont {Kakolee}, \citenamefont {Qiao}, \citenamefont {Macchi},
  \citenamefont {Cerchez}, \citenamefont {Doria}, \citenamefont {Geissler},
  \citenamefont {McKenna}, \citenamefont {Neely}, \citenamefont {Osterholz}
  \emph {et~al.}}]{kar2012ion}%
  \BibitemOpen
  \bibfield  {author} {\bibinfo {author} {\bibfnamefont {S.}~\bibnamefont
  {Kar}}, \bibinfo {author} {\bibfnamefont {K.}~\bibnamefont {Kakolee}},
  \bibinfo {author} {\bibfnamefont {B.}~\bibnamefont {Qiao}}, \bibinfo {author}
  {\bibfnamefont {A.}~\bibnamefont {Macchi}}, \bibinfo {author} {\bibfnamefont
  {M.}~\bibnamefont {Cerchez}}, \bibinfo {author} {\bibfnamefont
  {D.}~\bibnamefont {Doria}}, \bibinfo {author} {\bibfnamefont
  {M.}~\bibnamefont {Geissler}}, \bibinfo {author} {\bibfnamefont
  {P.}~\bibnamefont {McKenna}}, \bibinfo {author} {\bibfnamefont
  {D.}~\bibnamefont {Neely}}, \bibinfo {author} {\bibfnamefont
  {J.}~\bibnamefont {Osterholz}},  \emph {et~al.},\ }\bibfield  {title}
  {\enquote {\bibinfo {title} {Ion acceleration in multispecies targets driven
  by intense laser radiation pressure},}\ }\href@noop {} {\bibfield  {journal}
  {\bibinfo  {journal} {Physical Review Letters}\ }\textbf {\bibinfo {volume}
  {109}},\ \bibinfo {pages} {185006} (\bibinfo {year} {2012})}\BibitemShut
  {NoStop}%
\bibitem [{\citenamefont {Palaniyappan}\ \emph {et~al.}(2015)\citenamefont
  {Palaniyappan}, \citenamefont {Huang}, \citenamefont {Gautier}, \citenamefont
  {Hamilton}, \citenamefont {Santiago}, \citenamefont {Kreuzer}, \citenamefont
  {Sefkow}, \citenamefont {Shah},\ and\ \citenamefont
  {Fern{\'a}ndez}}]{palaniyappan2015efficient}%
  \BibitemOpen
  \bibfield  {author} {\bibinfo {author} {\bibfnamefont {S.}~\bibnamefont
  {Palaniyappan}}, \bibinfo {author} {\bibfnamefont {C.}~\bibnamefont {Huang}},
  \bibinfo {author} {\bibfnamefont {D.~C.}\ \bibnamefont {Gautier}}, \bibinfo
  {author} {\bibfnamefont {C.~E.}\ \bibnamefont {Hamilton}}, \bibinfo {author}
  {\bibfnamefont {M.~A.}\ \bibnamefont {Santiago}}, \bibinfo {author}
  {\bibfnamefont {C.}~\bibnamefont {Kreuzer}}, \bibinfo {author} {\bibfnamefont
  {A.~B.}\ \bibnamefont {Sefkow}}, \bibinfo {author} {\bibfnamefont {R.~C.}\
  \bibnamefont {Shah}}, \ and\ \bibinfo {author} {\bibfnamefont {J.~C.}\
  \bibnamefont {Fern{\'a}ndez}},\ }\bibfield  {title} {\enquote {\bibinfo
  {title} {Efficient quasi-monoenergetic ion beams from laser-driven
  relativistic plasmas},}\ }\href@noop {} {\bibfield  {journal} {\bibinfo
  {journal} {Nature communications}\ }\textbf {\bibinfo {volume} {6}},\
  \bibinfo {pages} {10170} (\bibinfo {year} {2015})}\BibitemShut {NoStop}%
\bibitem [{\citenamefont {Pak}\ \emph {et~al.}(2018)\citenamefont {Pak},
  \citenamefont {Kerr}, \citenamefont {Lemos}, \citenamefont {Link},
  \citenamefont {Patel}, \citenamefont {Albert}, \citenamefont {Divol},
  \citenamefont {Pollock}, \citenamefont {Haberberger}, \citenamefont {Froula}
  \emph {et~al.}}]{pak2018collisionless}%
  \BibitemOpen
  \bibfield  {author} {\bibinfo {author} {\bibfnamefont {A.}~\bibnamefont
  {Pak}}, \bibinfo {author} {\bibfnamefont {S.}~\bibnamefont {Kerr}}, \bibinfo
  {author} {\bibfnamefont {N.}~\bibnamefont {Lemos}}, \bibinfo {author}
  {\bibfnamefont {A.}~\bibnamefont {Link}}, \bibinfo {author} {\bibfnamefont
  {P.}~\bibnamefont {Patel}}, \bibinfo {author} {\bibfnamefont
  {F.}~\bibnamefont {Albert}}, \bibinfo {author} {\bibfnamefont
  {L.}~\bibnamefont {Divol}}, \bibinfo {author} {\bibfnamefont
  {B.}~\bibnamefont {Pollock}}, \bibinfo {author} {\bibfnamefont
  {D.}~\bibnamefont {Haberberger}}, \bibinfo {author} {\bibfnamefont
  {D.}~\bibnamefont {Froula}},  \emph {et~al.},\ }\bibfield  {title} {\enquote
  {\bibinfo {title} {Collisionless shock acceleration of narrow energy spread
  ion beams from mixed species plasmas using 1 $\mu$ m lasers},}\ }\href@noop
  {} {\bibfield  {journal} {\bibinfo  {journal} {Physical Review Accelerators
  and Beams}\ }\textbf {\bibinfo {volume} {21}},\ \bibinfo {pages} {103401}
  (\bibinfo {year} {2018})}\BibitemShut {NoStop}%
\bibitem [{\citenamefont {Scott}\ \emph {et~al.}(2018)\citenamefont {Scott},
  \citenamefont {Carroll}, \citenamefont {Astbury}, \citenamefont {Clarke},
  \citenamefont {Hernandez-Gomez}, \citenamefont {King}, \citenamefont {Alejo},
  \citenamefont {Arteaga}, \citenamefont {Dance}, \citenamefont {Higginson}
  \emph {et~al.}}]{scott2018dual}%
  \BibitemOpen
  \bibfield  {author} {\bibinfo {author} {\bibfnamefont {G.}~\bibnamefont
  {Scott}}, \bibinfo {author} {\bibfnamefont {D.}~\bibnamefont {Carroll}},
  \bibinfo {author} {\bibfnamefont {S.}~\bibnamefont {Astbury}}, \bibinfo
  {author} {\bibfnamefont {R.}~\bibnamefont {Clarke}}, \bibinfo {author}
  {\bibfnamefont {C.}~\bibnamefont {Hernandez-Gomez}}, \bibinfo {author}
  {\bibfnamefont {M.}~\bibnamefont {King}}, \bibinfo {author} {\bibfnamefont
  {A.}~\bibnamefont {Alejo}}, \bibinfo {author} {\bibfnamefont
  {I.}~\bibnamefont {Arteaga}}, \bibinfo {author} {\bibfnamefont
  {R.}~\bibnamefont {Dance}}, \bibinfo {author} {\bibfnamefont
  {A.}~\bibnamefont {Higginson}},  \emph {et~al.},\ }\bibfield  {title}
  {\enquote {\bibinfo {title} {Dual ion species plasma expansion from
  isotopically layered cryogenic targets},}\ }\href@noop {} {\bibfield
  {journal} {\bibinfo  {journal} {Physical review letters}\ }\textbf {\bibinfo
  {volume} {120}},\ \bibinfo {pages} {204801} (\bibinfo {year}
  {2018})}\BibitemShut {NoStop}%
\bibitem [{\citenamefont {Hilz}\ \emph {et~al.}(2018)\citenamefont {Hilz},
  \citenamefont {Ostermayr}, \citenamefont {Huebl}, \citenamefont {Bagnoud},
  \citenamefont {Borm}, \citenamefont {Bussmann}, \citenamefont {Gallei},
  \citenamefont {Gebhard}, \citenamefont {Haffa}, \citenamefont {Hartmann}
  \emph {et~al.}}]{hilz2018isolated}%
  \BibitemOpen
  \bibfield  {author} {\bibinfo {author} {\bibfnamefont {P.}~\bibnamefont
  {Hilz}}, \bibinfo {author} {\bibfnamefont {T.}~\bibnamefont {Ostermayr}},
  \bibinfo {author} {\bibfnamefont {A.}~\bibnamefont {Huebl}}, \bibinfo
  {author} {\bibfnamefont {V.}~\bibnamefont {Bagnoud}}, \bibinfo {author}
  {\bibfnamefont {B.}~\bibnamefont {Borm}}, \bibinfo {author} {\bibfnamefont
  {M.}~\bibnamefont {Bussmann}}, \bibinfo {author} {\bibfnamefont
  {M.}~\bibnamefont {Gallei}}, \bibinfo {author} {\bibfnamefont
  {J.}~\bibnamefont {Gebhard}}, \bibinfo {author} {\bibfnamefont
  {D.}~\bibnamefont {Haffa}}, \bibinfo {author} {\bibfnamefont
  {J.}~\bibnamefont {Hartmann}},  \emph {et~al.},\ }\bibfield  {title}
  {\enquote {\bibinfo {title} {Isolated proton bunch acceleration by a petawatt
  laser pulse},}\ }\href@noop {} {\bibfield  {journal} {\bibinfo  {journal}
  {Nature communications}\ }\textbf {\bibinfo {volume} {9}},\ \bibinfo {pages}
  {423} (\bibinfo {year} {2018})}\BibitemShut {NoStop}%
\bibitem [{\citenamefont {Li}\ \emph {et~al.}(2019)\citenamefont {Li},
  \citenamefont {Forestier-Colleoni}, \citenamefont {Bailly-Grandvaux},
  \citenamefont {McGuffey}, \citenamefont {Arefiev}, \citenamefont {Bulanov},
  \citenamefont {Peebles}, \citenamefont {Krauland}, \citenamefont {Hussein},
  \citenamefont {Batson} \emph {et~al.}}]{li2019laser}%
  \BibitemOpen
  \bibfield  {author} {\bibinfo {author} {\bibfnamefont {J.}~\bibnamefont
  {Li}}, \bibinfo {author} {\bibfnamefont {P.}~\bibnamefont
  {Forestier-Colleoni}}, \bibinfo {author} {\bibfnamefont {M.}~\bibnamefont
  {Bailly-Grandvaux}}, \bibinfo {author} {\bibfnamefont {C.}~\bibnamefont
  {McGuffey}}, \bibinfo {author} {\bibfnamefont {A.}~\bibnamefont {Arefiev}},
  \bibinfo {author} {\bibfnamefont {S.}~\bibnamefont {Bulanov}}, \bibinfo
  {author} {\bibfnamefont {J.}~\bibnamefont {Peebles}}, \bibinfo {author}
  {\bibfnamefont {C.}~\bibnamefont {Krauland}}, \bibinfo {author}
  {\bibfnamefont {A.}~\bibnamefont {Hussein}}, \bibinfo {author} {\bibfnamefont
  {T.}~\bibnamefont {Batson}},  \emph {et~al.},\ }\bibfield  {title} {\enquote
  {\bibinfo {title} {Laser-driven acceleration of quasi-monoenergetic,
  near-collimated titanium ions via a transparency-enhanced acceleration
  scheme},}\ }\href@noop {} {\bibfield  {journal} {\bibinfo  {journal} {New
  Journal of Physics}\ }\textbf {\bibinfo {volume} {21}},\ \bibinfo {pages}
  {103005} (\bibinfo {year} {2019})}\BibitemShut {NoStop}%
\bibitem [{\citenamefont {Bagchi}\ \emph {et~al.}(2021)\citenamefont {Bagchi},
  \citenamefont {Tayyab}, \citenamefont {Pasley}, \citenamefont {Robinson},
  \citenamefont {Nayak},\ and\ \citenamefont {Chakera}}]{bagchi2021quasi}%
  \BibitemOpen
  \bibfield  {author} {\bibinfo {author} {\bibfnamefont {S.}~\bibnamefont
  {Bagchi}}, \bibinfo {author} {\bibfnamefont {M.}~\bibnamefont {Tayyab}},
  \bibinfo {author} {\bibfnamefont {J.}~\bibnamefont {Pasley}}, \bibinfo
  {author} {\bibfnamefont {A.~P.}\ \bibnamefont {Robinson}}, \bibinfo {author}
  {\bibfnamefont {M.}~\bibnamefont {Nayak}}, \ and\ \bibinfo {author}
  {\bibfnamefont {J.~A.}\ \bibnamefont {Chakera}},\ }\bibfield  {title}
  {\enquote {\bibinfo {title} {Quasi mono-energetic heavy ion acceleration from
  layered targets},}\ }\href@noop {} {\bibfield  {journal} {\bibinfo  {journal}
  {Physics of Plasmas}\ }\textbf {\bibinfo {volume} {28}} (\bibinfo {year}
  {2021})}\BibitemShut {NoStop}%
\bibitem [{\citenamefont {Ahmed}\ \emph {et~al.}(2021)\citenamefont {Ahmed},
  \citenamefont {Hadjisolomou}, \citenamefont {Naughton}, \citenamefont
  {Alejo}, \citenamefont {Brauckmann}, \citenamefont {Cantono}, \citenamefont
  {Ferguson}, \citenamefont {Cerchez}, \citenamefont {Doria}, \citenamefont
  {Green} \emph {et~al.}}]{ahmed2021high}%
  \BibitemOpen
  \bibfield  {author} {\bibinfo {author} {\bibfnamefont {H.}~\bibnamefont
  {Ahmed}}, \bibinfo {author} {\bibfnamefont {P.}~\bibnamefont {Hadjisolomou}},
  \bibinfo {author} {\bibfnamefont {K.}~\bibnamefont {Naughton}}, \bibinfo
  {author} {\bibfnamefont {A.}~\bibnamefont {Alejo}}, \bibinfo {author}
  {\bibfnamefont {S.}~\bibnamefont {Brauckmann}}, \bibinfo {author}
  {\bibfnamefont {G.}~\bibnamefont {Cantono}}, \bibinfo {author} {\bibfnamefont
  {S.}~\bibnamefont {Ferguson}}, \bibinfo {author} {\bibfnamefont
  {M.}~\bibnamefont {Cerchez}}, \bibinfo {author} {\bibfnamefont
  {D.}~\bibnamefont {Doria}}, \bibinfo {author} {\bibfnamefont
  {J.}~\bibnamefont {Green}},  \emph {et~al.},\ }\bibfield  {title} {\enquote
  {\bibinfo {title} {High energy implementation of coil-target scheme for
  guided re-acceleration of laser-driven protons},}\ }\href@noop {} {\bibfield
  {journal} {\bibinfo  {journal} {Scientific Reports}\ }\textbf {\bibinfo
  {volume} {11}},\ \bibinfo {pages} {699} (\bibinfo {year} {2021})}\BibitemShut
  {NoStop}%
\bibitem [{\citenamefont {Wilks}\ \emph {et~al.}(2001)\citenamefont {Wilks},
  \citenamefont {Langdon}, \citenamefont {Cowan}, \citenamefont {Roth},
  \citenamefont {Singh}, \citenamefont {Hatchett}, \citenamefont {Key},
  \citenamefont {Pennington}, \citenamefont {MacKinnon},\ and\ \citenamefont
  {Snavely}}]{wilks2001energetic}%
  \BibitemOpen
  \bibfield  {author} {\bibinfo {author} {\bibfnamefont {S.}~\bibnamefont
  {Wilks}}, \bibinfo {author} {\bibfnamefont {A.}~\bibnamefont {Langdon}},
  \bibinfo {author} {\bibfnamefont {T.}~\bibnamefont {Cowan}}, \bibinfo
  {author} {\bibfnamefont {M.}~\bibnamefont {Roth}}, \bibinfo {author}
  {\bibfnamefont {M.}~\bibnamefont {Singh}}, \bibinfo {author} {\bibfnamefont
  {S.}~\bibnamefont {Hatchett}}, \bibinfo {author} {\bibfnamefont
  {M.}~\bibnamefont {Key}}, \bibinfo {author} {\bibfnamefont {D.}~\bibnamefont
  {Pennington}}, \bibinfo {author} {\bibfnamefont {A.}~\bibnamefont
  {MacKinnon}}, \ and\ \bibinfo {author} {\bibfnamefont {R.}~\bibnamefont
  {Snavely}},\ }\bibfield  {title} {\enquote {\bibinfo {title} {Energetic
  proton generation in ultra-intense laser--solid interactions},}\ }\href@noop
  {} {\bibfield  {journal} {\bibinfo  {journal} {Physics of plasmas}\ }\textbf
  {\bibinfo {volume} {8}},\ \bibinfo {pages} {542--549} (\bibinfo {year}
  {2001})}\BibitemShut {NoStop}%
\bibitem [{\citenamefont {Passoni}, \citenamefont {Bertagna},\ and\
  \citenamefont {Zani}(2010)}]{passoni2010target}%
  \BibitemOpen
  \bibfield  {author} {\bibinfo {author} {\bibfnamefont {M.}~\bibnamefont
  {Passoni}}, \bibinfo {author} {\bibfnamefont {L.}~\bibnamefont {Bertagna}}, \
  and\ \bibinfo {author} {\bibfnamefont {A.}~\bibnamefont {Zani}},\ }\bibfield
  {title} {\enquote {\bibinfo {title} {Target normal sheath acceleration:
  theory, comparison with experiments and future perspectives},}\ }\href@noop
  {} {\bibfield  {journal} {\bibinfo  {journal} {New Journal of Physics}\
  }\textbf {\bibinfo {volume} {12}},\ \bibinfo {pages} {045012} (\bibinfo
  {year} {2010})}\BibitemShut {NoStop}%
\bibitem [{\citenamefont {Esirkepov}\ \emph {et~al.}(2004)\citenamefont
  {Esirkepov}, \citenamefont {Borghesi}, \citenamefont {Bulanov}, \citenamefont
  {Mourou},\ and\ \citenamefont {Tajima}}]{esirkepov2004highly}%
  \BibitemOpen
  \bibfield  {author} {\bibinfo {author} {\bibfnamefont {T.}~\bibnamefont
  {Esirkepov}}, \bibinfo {author} {\bibfnamefont {M.}~\bibnamefont {Borghesi}},
  \bibinfo {author} {\bibfnamefont {S.}~\bibnamefont {Bulanov}}, \bibinfo
  {author} {\bibfnamefont {G.}~\bibnamefont {Mourou}}, \ and\ \bibinfo {author}
  {\bibfnamefont {T.}~\bibnamefont {Tajima}},\ }\bibfield  {title} {\enquote
  {\bibinfo {title} {Highly efficient relativistic-ion generation in the
  laser-piston regime},}\ }\href@noop {} {\bibfield  {journal} {\bibinfo
  {journal} {Physical review letters}\ }\textbf {\bibinfo {volume} {92}},\
  \bibinfo {pages} {175003} (\bibinfo {year} {2004})}\BibitemShut {NoStop}%
\bibitem [{\citenamefont {Silva}\ \emph {et~al.}(2004)\citenamefont {Silva},
  \citenamefont {Marti}, \citenamefont {Davies}, \citenamefont {Fonseca},
  \citenamefont {Ren}, \citenamefont {Tsung},\ and\ \citenamefont
  {Mori}}]{silva2004proton}%
  \BibitemOpen
  \bibfield  {author} {\bibinfo {author} {\bibfnamefont {L.~O.}\ \bibnamefont
  {Silva}}, \bibinfo {author} {\bibfnamefont {M.}~\bibnamefont {Marti}},
  \bibinfo {author} {\bibfnamefont {J.~R.}\ \bibnamefont {Davies}}, \bibinfo
  {author} {\bibfnamefont {R.~A.}\ \bibnamefont {Fonseca}}, \bibinfo {author}
  {\bibfnamefont {C.}~\bibnamefont {Ren}}, \bibinfo {author} {\bibfnamefont
  {F.~S.}\ \bibnamefont {Tsung}}, \ and\ \bibinfo {author} {\bibfnamefont
  {W.~B.}\ \bibnamefont {Mori}},\ }\bibfield  {title} {\enquote {\bibinfo
  {title} {Proton shock acceleration in laser-plasma interactions},}\
  }\href@noop {} {\bibfield  {journal} {\bibinfo  {journal} {Physical Review
  Letters}\ }\textbf {\bibinfo {volume} {92}},\ \bibinfo {pages} {015002}
  (\bibinfo {year} {2004})}\BibitemShut {NoStop}%
\bibitem [{\citenamefont {Prencipe}\ \emph {et~al.}(2017)\citenamefont
  {Prencipe}, \citenamefont {Fuchs}, \citenamefont {Pascarelli}, \citenamefont
  {Schumacher}, \citenamefont {Stephens}, \citenamefont {Alexander},
  \citenamefont {Briggs}, \citenamefont {B{\"u}scher}, \citenamefont
  {Cernaianu}, \citenamefont {Choukourov} \emph
  {et~al.}}]{prencipe2017targets}%
  \BibitemOpen
  \bibfield  {author} {\bibinfo {author} {\bibfnamefont {I.}~\bibnamefont
  {Prencipe}}, \bibinfo {author} {\bibfnamefont {J.}~\bibnamefont {Fuchs}},
  \bibinfo {author} {\bibfnamefont {S.}~\bibnamefont {Pascarelli}}, \bibinfo
  {author} {\bibfnamefont {D.}~\bibnamefont {Schumacher}}, \bibinfo {author}
  {\bibfnamefont {R.}~\bibnamefont {Stephens}}, \bibinfo {author}
  {\bibfnamefont {N.}~\bibnamefont {Alexander}}, \bibinfo {author}
  {\bibfnamefont {R.}~\bibnamefont {Briggs}}, \bibinfo {author} {\bibfnamefont
  {M.}~\bibnamefont {B{\"u}scher}}, \bibinfo {author} {\bibfnamefont
  {M.}~\bibnamefont {Cernaianu}}, \bibinfo {author} {\bibfnamefont
  {A.}~\bibnamefont {Choukourov}},  \emph {et~al.},\ }\bibfield  {title}
  {\enquote {\bibinfo {title} {Targets for high repetition rate laser
  facilities: needs, challenges and perspectives},}\ }\href@noop {} {\bibfield
  {journal} {\bibinfo  {journal} {High Power Laser Science and Engineering}\
  }\textbf {\bibinfo {volume} {5}},\ \bibinfo {pages} {e17} (\bibinfo {year}
  {2017})}\BibitemShut {NoStop}%
\bibitem [{\citenamefont {Hou}\ \emph {et~al.}(2011)\citenamefont {Hou},
  \citenamefont {Nees}, \citenamefont {He}, \citenamefont {Petrov},
  \citenamefont {Davis}, \citenamefont {Easter}, \citenamefont {Thomas},\ and\
  \citenamefont {Krushelnick}}]{hou2011laser}%
  \BibitemOpen
  \bibfield  {author} {\bibinfo {author} {\bibfnamefont {B.}~\bibnamefont
  {Hou}}, \bibinfo {author} {\bibfnamefont {J.~A.}\ \bibnamefont {Nees}},
  \bibinfo {author} {\bibfnamefont {Z.}~\bibnamefont {He}}, \bibinfo {author}
  {\bibfnamefont {G.}~\bibnamefont {Petrov}}, \bibinfo {author} {\bibfnamefont
  {J.}~\bibnamefont {Davis}}, \bibinfo {author} {\bibfnamefont {J.~H.}\
  \bibnamefont {Easter}}, \bibinfo {author} {\bibfnamefont {A.~G.}\
  \bibnamefont {Thomas}}, \ and\ \bibinfo {author} {\bibfnamefont {K.~M.}\
  \bibnamefont {Krushelnick}},\ }\bibfield  {title} {\enquote {\bibinfo {title}
  {Laser-ion acceleration through controlled surface contamination},}\
  }\href@noop {} {\bibfield  {journal} {\bibinfo  {journal} {Physics of
  Plasmas}\ }\textbf {\bibinfo {volume} {18}} (\bibinfo {year}
  {2011})}\BibitemShut {NoStop}%
\bibitem [{\citenamefont {Alejo}\ \emph {et~al.}(2016)\citenamefont {Alejo},
  \citenamefont {Kar}, \citenamefont {Tebartz}, \citenamefont {Ahmed},
  \citenamefont {Astbury}, \citenamefont {Carroll}, \citenamefont {Ding},
  \citenamefont {Doria}, \citenamefont {Higginson}, \citenamefont {McKenna}
  \emph {et~al.}}]{alejo2016high}%
  \BibitemOpen
  \bibfield  {author} {\bibinfo {author} {\bibfnamefont {A.}~\bibnamefont
  {Alejo}}, \bibinfo {author} {\bibfnamefont {S.}~\bibnamefont {Kar}}, \bibinfo
  {author} {\bibfnamefont {A.}~\bibnamefont {Tebartz}}, \bibinfo {author}
  {\bibfnamefont {H.}~\bibnamefont {Ahmed}}, \bibinfo {author} {\bibfnamefont
  {S.}~\bibnamefont {Astbury}}, \bibinfo {author} {\bibfnamefont
  {D.}~\bibnamefont {Carroll}}, \bibinfo {author} {\bibfnamefont
  {J.}~\bibnamefont {Ding}}, \bibinfo {author} {\bibfnamefont {D.}~\bibnamefont
  {Doria}}, \bibinfo {author} {\bibfnamefont {A.}~\bibnamefont {Higginson}},
  \bibinfo {author} {\bibfnamefont {P.}~\bibnamefont {McKenna}},  \emph
  {et~al.},\ }\bibfield  {title} {\enquote {\bibinfo {title} {High resolution
  thomson parabola spectrometer for full spectral capture of multi-species ion
  beams},}\ }\href@noop {} {\bibfield  {journal} {\bibinfo  {journal} {Review
  of Scientific Instruments}\ }\textbf {\bibinfo {volume} {87}} (\bibinfo
  {year} {2016})}\BibitemShut {NoStop}%
\bibitem [{\citenamefont {Tosaki}\ \emph {et~al.}(2017)\citenamefont {Tosaki},
  \citenamefont {Yogo}, \citenamefont {Koga}, \citenamefont {Okamoto},
  \citenamefont {Shokita}, \citenamefont {Morace}, \citenamefont {Arikawa},
  \citenamefont {Fujioka}, \citenamefont {Nakai}, \citenamefont {Shiraga} \emph
  {et~al.}}]{tosaki2017evaluation}%
  \BibitemOpen
  \bibfield  {author} {\bibinfo {author} {\bibfnamefont {S.}~\bibnamefont
  {Tosaki}}, \bibinfo {author} {\bibfnamefont {A.}~\bibnamefont {Yogo}},
  \bibinfo {author} {\bibfnamefont {K.}~\bibnamefont {Koga}}, \bibinfo {author}
  {\bibfnamefont {K.}~\bibnamefont {Okamoto}}, \bibinfo {author} {\bibfnamefont
  {S.}~\bibnamefont {Shokita}}, \bibinfo {author} {\bibfnamefont
  {A.}~\bibnamefont {Morace}}, \bibinfo {author} {\bibfnamefont
  {Y.}~\bibnamefont {Arikawa}}, \bibinfo {author} {\bibfnamefont
  {S.}~\bibnamefont {Fujioka}}, \bibinfo {author} {\bibfnamefont
  {M.}~\bibnamefont {Nakai}}, \bibinfo {author} {\bibfnamefont
  {H.}~\bibnamefont {Shiraga}},  \emph {et~al.},\ }\bibfield  {title} {\enquote
  {\bibinfo {title} {Evaluation of laser-driven ion energies for fusion
  fast-ignition research},}\ }\href@noop {} {\bibfield  {journal} {\bibinfo
  {journal} {Progress of Theoretical and Experimental Physics}\ }\textbf
  {\bibinfo {volume} {2017}},\ \bibinfo {pages} {103J01} (\bibinfo {year}
  {2017})}\BibitemShut {NoStop}%
\bibitem [{\citenamefont {Takagi}\ \emph {et~al.}(1991)\citenamefont {Takagi},
  \citenamefont {Norimatsu}, \citenamefont {Yamanaka},\ and\ \citenamefont
  {Nakai}}]{takagi1991development}%
  \BibitemOpen
  \bibfield  {author} {\bibinfo {author} {\bibfnamefont {M.}~\bibnamefont
  {Takagi}}, \bibinfo {author} {\bibfnamefont {T.}~\bibnamefont {Norimatsu}},
  \bibinfo {author} {\bibfnamefont {T.}~\bibnamefont {Yamanaka}}, \ and\
  \bibinfo {author} {\bibfnamefont {S.}~\bibnamefont {Nakai}},\ }\bibfield
  {title} {\enquote {\bibinfo {title} {Development of deuterated polystyrene
  shells for laser fusion by means of a density-matched emulsion method},}\
  }\href@noop {} {\bibfield  {journal} {\bibinfo  {journal} {Journal of Vacuum
  Science \& Technology A: Vacuum, Surfaces, and Films}\ }\textbf {\bibinfo
  {volume} {9}},\ \bibinfo {pages} {2145--2148} (\bibinfo {year}
  {1991})}\BibitemShut {NoStop}%
\bibitem [{\citenamefont {Iwasa}\ \emph {et~al.}(2018)\citenamefont {Iwasa},
  \citenamefont {Yamanoi}, \citenamefont {Kaneyasu},\ and\ \citenamefont
  {Norimatsu}}]{iwasa2018controlled}%
  \BibitemOpen
  \bibfield  {author} {\bibinfo {author} {\bibfnamefont {Y.}~\bibnamefont
  {Iwasa}}, \bibinfo {author} {\bibfnamefont {K.}~\bibnamefont {Yamanoi}},
  \bibinfo {author} {\bibfnamefont {Y.}~\bibnamefont {Kaneyasu}}, \ and\
  \bibinfo {author} {\bibfnamefont {T.}~\bibnamefont {Norimatsu}},\ }\bibfield
  {title} {\enquote {\bibinfo {title} {Controlled generation of double
  emulsions for laser fusion target fabrication using a glass capillary
  microfluidic device},}\ }\href@noop {} {\bibfield  {journal} {\bibinfo
  {journal} {Fusion Science and Technology}\ }\textbf {\bibinfo {volume}
  {73}},\ \bibinfo {pages} {258--264} (\bibinfo {year} {2018})}\BibitemShut
  {NoStop}%
\bibitem [{\citenamefont {Davis}(2002)}]{davis2002ordered}%
  \BibitemOpen
  \bibfield  {author} {\bibinfo {author} {\bibfnamefont {M.~E.}\ \bibnamefont
  {Davis}},\ }\bibfield  {title} {\enquote {\bibinfo {title} {Ordered porous
  materials for emerging applications},}\ }\href@noop {} {\bibfield  {journal}
  {\bibinfo  {journal} {Nature}\ }\textbf {\bibinfo {volume} {417}},\ \bibinfo
  {pages} {813--821} (\bibinfo {year} {2002})}\BibitemShut {NoStop}%
\bibitem [{\citenamefont {Kawanaka}\ \emph {et~al.}(2008)\citenamefont
  {Kawanaka}, \citenamefont {Miyanaga}, \citenamefont {Azechi}, \citenamefont
  {Kanabe}, \citenamefont {Jitsuno}, \citenamefont {Kondo}, \citenamefont
  {Fujimoto}, \citenamefont {Morio}, \citenamefont {Matsuo}, \citenamefont
  {Kawakami} \emph {et~al.}}]{kawanaka20083}%
  \BibitemOpen
  \bibfield  {author} {\bibinfo {author} {\bibfnamefont {J.}~\bibnamefont
  {Kawanaka}}, \bibinfo {author} {\bibfnamefont {N.}~\bibnamefont {Miyanaga}},
  \bibinfo {author} {\bibfnamefont {H.}~\bibnamefont {Azechi}}, \bibinfo
  {author} {\bibfnamefont {T.}~\bibnamefont {Kanabe}}, \bibinfo {author}
  {\bibfnamefont {T.}~\bibnamefont {Jitsuno}}, \bibinfo {author} {\bibfnamefont
  {K.}~\bibnamefont {Kondo}}, \bibinfo {author} {\bibfnamefont
  {Y.}~\bibnamefont {Fujimoto}}, \bibinfo {author} {\bibfnamefont
  {N.}~\bibnamefont {Morio}}, \bibinfo {author} {\bibfnamefont
  {S.}~\bibnamefont {Matsuo}}, \bibinfo {author} {\bibfnamefont
  {Y.}~\bibnamefont {Kawakami}},  \emph {et~al.},\ }\bibfield  {title}
  {\enquote {\bibinfo {title} {3.1-kj chirped-pulse power amplification in the
  lfex laser},}\ }in\ \href@noop {} {\emph {\bibinfo {booktitle} {Journal of
  Physics: Conference Series}}},\ Vol.\ \bibinfo {volume} {112}\ (\bibinfo
  {organization} {IOP Publishing},\ \bibinfo {year} {2008})\ p.\ \bibinfo
  {pages} {032006}\BibitemShut {NoStop}%
\bibitem [{\citenamefont {Golovin}\ \emph {et~al.}(2021)\citenamefont
  {Golovin}, \citenamefont {Mirfayzi}, \citenamefont {Shokita}, \citenamefont
  {Abe}, \citenamefont {Lan}, \citenamefont {Arikawa}, \citenamefont {Morace},
  \citenamefont {Pikuz},\ and\ \citenamefont {Yogo}}]{golovin2021calibration}%
  \BibitemOpen
  \bibfield  {author} {\bibinfo {author} {\bibfnamefont {D.}~\bibnamefont
  {Golovin}}, \bibinfo {author} {\bibfnamefont {S.}~\bibnamefont {Mirfayzi}},
  \bibinfo {author} {\bibfnamefont {S.}~\bibnamefont {Shokita}}, \bibinfo
  {author} {\bibfnamefont {Y.}~\bibnamefont {Abe}}, \bibinfo {author}
  {\bibfnamefont {Z.}~\bibnamefont {Lan}}, \bibinfo {author} {\bibfnamefont
  {Y.}~\bibnamefont {Arikawa}}, \bibinfo {author} {\bibfnamefont
  {A.}~\bibnamefont {Morace}}, \bibinfo {author} {\bibfnamefont
  {T.}~\bibnamefont {Pikuz}}, \ and\ \bibinfo {author} {\bibfnamefont
  {A.}~\bibnamefont {Yogo}},\ }\bibfield  {title} {\enquote {\bibinfo {title}
  {Calibration of imaging plates sensitivity to high energy photons and ions
  for laser-plasma interaction sources},}\ }\href@noop {} {\bibfield  {journal}
  {\bibinfo  {journal} {Journal of Instrumentation}\ }\textbf {\bibinfo
  {volume} {16}},\ \bibinfo {pages} {T02005} (\bibinfo {year}
  {2021})}\BibitemShut {NoStop}%
\bibitem [{\citenamefont {Alfonso}, \citenamefont {Jaquez},\ and\ \citenamefont
  {Nikroo}(2006)}]{alfonso2006using}%
  \BibitemOpen
  \bibfield  {author} {\bibinfo {author} {\bibfnamefont {E.}~\bibnamefont
  {Alfonso}}, \bibinfo {author} {\bibfnamefont {J.}~\bibnamefont {Jaquez}}, \
  and\ \bibinfo {author} {\bibfnamefont {A.}~\bibnamefont {Nikroo}},\
  }\bibfield  {title} {\enquote {\bibinfo {title} {Using mass spectrometry to
  characterize permeation half-life of icf targets},}\ }\href@noop {}
  {\bibfield  {journal} {\bibinfo  {journal} {Fusion science and technology}\
  }\textbf {\bibinfo {volume} {49}},\ \bibinfo {pages} {773--777} (\bibinfo
  {year} {2006})}\BibitemShut {NoStop}%
\bibitem [{\citenamefont {Bennett}\ \emph {et~al.}(2017)\citenamefont
  {Bennett}, \citenamefont {Brady}, \citenamefont {Schmitz}, \citenamefont
  {Ridgers}, \citenamefont {Arber}, \citenamefont {Evans},\ and\ \citenamefont
  {Bell}}]{bennett2017users}%
  \BibitemOpen
  \bibfield  {author} {\bibinfo {author} {\bibfnamefont {K.}~\bibnamefont
  {Bennett}}, \bibinfo {author} {\bibfnamefont {C.}~\bibnamefont {Brady}},
  \bibinfo {author} {\bibfnamefont {H.}~\bibnamefont {Schmitz}}, \bibinfo
  {author} {\bibfnamefont {C.}~\bibnamefont {Ridgers}}, \bibinfo {author}
  {\bibfnamefont {T.}~\bibnamefont {Arber}}, \bibinfo {author} {\bibfnamefont
  {R.}~\bibnamefont {Evans}}, \ and\ \bibinfo {author} {\bibfnamefont
  {T.}~\bibnamefont {Bell}},\ }\bibfield  {title} {\enquote {\bibinfo {title}
  {Users manual for the epoch pic codes},}\ }\href@noop {} {\bibfield
  {journal} {\bibinfo  {journal} {University of Warwick}\ } (\bibinfo {year}
  {2017})}\BibitemShut {NoStop}%
\bibitem [{\citenamefont {Iwata}\ \emph {et~al.}(2021)\citenamefont {Iwata},
  \citenamefont {Kemp}, \citenamefont {Wilks}, \citenamefont {Mima},
  \citenamefont {Mariscal}, \citenamefont {Ma},\ and\ \citenamefont
  {Sentoku}}]{iwata2021lateral}%
  \BibitemOpen
  \bibfield  {author} {\bibinfo {author} {\bibfnamefont {N.}~\bibnamefont
  {Iwata}}, \bibinfo {author} {\bibfnamefont {A.}~\bibnamefont {Kemp}},
  \bibinfo {author} {\bibfnamefont {S.}~\bibnamefont {Wilks}}, \bibinfo
  {author} {\bibfnamefont {K.}~\bibnamefont {Mima}}, \bibinfo {author}
  {\bibfnamefont {D.}~\bibnamefont {Mariscal}}, \bibinfo {author}
  {\bibfnamefont {T.}~\bibnamefont {Ma}}, \ and\ \bibinfo {author}
  {\bibfnamefont {Y.}~\bibnamefont {Sentoku}},\ }\bibfield  {title} {\enquote
  {\bibinfo {title} {Lateral confinement of fast electrons and its impact on
  laser ion acceleration},}\ }\href@noop {} {\bibfield  {journal} {\bibinfo
  {journal} {Physical Review Research}\ }\textbf {\bibinfo {volume} {3}},\
  \bibinfo {pages} {023193} (\bibinfo {year} {2021})}\BibitemShut {NoStop}%
\bibitem [{\citenamefont {Schwemmlein}\ \emph {et~al.}(2022)\citenamefont
  {Schwemmlein}, \citenamefont {Stoeckl}, \citenamefont {Forrest},
  \citenamefont {Shmayda}, \citenamefont {Regan},\ and\ \citenamefont
  {Schr{\"o}der}}]{schwemmlein2022first}%
  \BibitemOpen
  \bibfield  {author} {\bibinfo {author} {\bibfnamefont {A.}~\bibnamefont
  {Schwemmlein}}, \bibinfo {author} {\bibfnamefont {C.}~\bibnamefont
  {Stoeckl}}, \bibinfo {author} {\bibfnamefont {C.}~\bibnamefont {Forrest}},
  \bibinfo {author} {\bibfnamefont {W.}~\bibnamefont {Shmayda}}, \bibinfo
  {author} {\bibfnamefont {S.}~\bibnamefont {Regan}}, \ and\ \bibinfo {author}
  {\bibfnamefont {W.}~\bibnamefont {Schr{\"o}der}},\ }\bibfield  {title}
  {\enquote {\bibinfo {title} {First demonstration of a triton beam using
  target normal sheath acceleration},}\ }\href@noop {} {\bibfield  {journal}
  {\bibinfo  {journal} {Nuclear Instruments and Methods in Physics Research
  Section B: Beam Interactions with Materials and Atoms}\ }\textbf {\bibinfo
  {volume} {522}},\ \bibinfo {pages} {27--31} (\bibinfo {year}
  {2022})}\BibitemShut {NoStop}%
\end{thebibliography}%

\end{document}